\documentclass[traditabstract]{aa}
\usepackage{verbatim} 
\usepackage{amsmath}
\usepackage{amssymb}
\usepackage{graphicx}
\usepackage{booktabs}
\usepackage{xspace}
\usepackage{siunitx}
\usepackage{mathrsfs}
\usepackage{placeins}
\usepackage{subcaption}

\makeatletter

\usepackage{txfonts}
\usepackage{enumitem}
\usepackage{natbib,twoopt}
\usepackage[breaklinks=true,colorlinks=true,linkcolor=blue,urlcolor=blue,citecolor=blue]{hyperref} 
\bibpunct{(}{)}{;}{a}{}{,}        
\makeatletter
  \newcommandtwoopt{\citeads}[3][][]{\href{http://adsabs.harvard.edu/abs/#3}%
    {\def\hyper@linkstart##1##2{}%
     \let\hyper@linkend\@empty\citealp[#1][#2]{#3}}}
  \newcommandtwoopt{\citepads}[3][][]{\href{http://adsabs.harvard.edu/abs/#3}%
    {\def\hyper@linkstart##1##2{}%
     \let\hyper@linkend\@empty\citep[#1][#2]{#3}}}
  \newcommandtwoopt{\citetads}[3][][]{\href{http://adsabs.harvard.edu/abs/#3}%
    {\def\hyper@linkstart##1##2{}%
     \let\hyper@linkend\@empty\citet[#1][#2]{#3}}}
  \newcommandtwoopt{\citeyearads}[3][][]%
    {\href{http://adsabs.harvard.edu/abs/#3}
    {\def\hyper@linkstart##1##2{}%
     \let\hyper@linkend\@empty\citeyear[#1][#2]{#3}}}
\makeatother

\newcommand{\msun}{{\rm M}_{\odot}}
\newcommand{\lsun}{{\rm L}_{\odot}}
\newcommand{\rsun}{{\rm R}_{\odot}}

\newcommand{\eg}{e.g.\@\xspace}

\newcommand{\ie}{i.e.\@\xspace}

\titlerunning{Asteroseismic predictions for a massive MS merger product}
\authorrunning{J.~Henneco et al.}

\makeatother

\begin{document}

\title{Asteroseismic predictions for a massive main-sequence merger product}

\author{
J.~Henneco\inst{\ref{HITS},\,\ref{UNIHD}}\thanks{jan.henneco@protonmail.com}, F.~R.~N.~~Schneider\inst{\ref{HITS},\,\ref{ZAHARI}},  M.~Heller\inst{\ref{HITS},\,\ref{UNIHD}}, S.~Hekker\inst{\ref{HITS},\,\ref{ZAHLSW}},  \and C.~Aerts\inst{\ref{IVS},\,\ref{RUN},\,\ref{MPIA}}
}

\institute{Heidelberg Institute for Theoretical Studies, Schloss-Wolfsbrunnenweg 35, 69118 Heidelberg, Germany\label{HITS}
\and Universit\"at Heidelberg, Department of Physics and Astronomy, Im Neuenheimer Feld 226, 69120 Heidelberg, Germany\label{UNIHD}
\and Zentrum f\"ur Astronomie, Astronomisches Rechen-Institut (ZAH/ARI), Heidelberg University,  M\"{o}nchhofstr. 12-14, 69120 Heidelberg, Germany\label{ZAHARI}
\and Zentrum f\"ur Astronomie, Landessternwarte (ZAH/LSW), Heidelberg University, K\"{o}nigstuhl 12, 69117 Heidelberg, Germany\label{ZAHLSW}
\and Institute of Astronomy, KU Leuven, Celestijnenlaan 200D, 3001 Leuven, Belgium\label{IVS}
\and Department of Astrophysics, IMAPP, Radboud University Nijmegen, PO Box 9010, 6500 GL, Nijmegen, The Netherlands\label{RUN}
\and Max Planck Institute for Astronomy, K\"{o}nigstuhl 17, 69117, Heidelberg, Germany\label{MPIA}}

\date{Received 20 December 2024 / Accepted 7 April 2025}

\abstract{The products of stellar mergers between two massive main-sequence stars appear as seemingly normal main-sequence stars after a phase of thermal relaxation, if not for certain peculiarities. These peculiarities, such as strong magnetic fields, chemically enriched surfaces, rejuvenated cores, and masses above the main-sequence turnoff mass, have been proposed to indicate merger or mass accretion origins. Since these peculiarities are not limited to the merger product's surface, we use asteroseismology to predict how the differences in the internal structure of a merger product and a genuine single star manifest via properties of non-radial stellar pulsations. We use the result of a 3D (magneto)hydrodynamic simulation of a stellar merger between a 9 and an $8\,\msun$ main-sequence star, which was mapped to 1D and evolved through the main sequence. We compare the predicted pressure and gravity modes for the merger product model with those predicted for a corresponding genuine single-star model. The pressure-mode frequencies are consistently lower for the merger product than for the genuine single star, and the differences between them are more than a thousand times larger than the current best observational uncertainties for measured mode frequencies of this kind. Even though the absolute differences in gravity-mode period spacings vary in value and sign throughout the main-sequence life of both stars, they, too, are larger than the current best observational uncertainties for such long-period modes. This, combined with additional variability in the merger product's period spacing patterns, shows the potential of identifying merger products in future-forward modelling. We also attempt to replicate the merger product's structure using three widely applied 1D merger prescriptions and repeat the asteroseismic analysis. Although none of the 1D prescriptions reproduces the entire merger product's structure, we conclude that the prescription with shock heating shows the highest potential, provided that it can be calibrated on binary-evolution-driven 3D merger simulations. Our work focuses on a particular kind of massive main-sequence merger and should be expanded to encompass the various possible merger product structures predicted to exist in the Universe.}


\maketitle

\section{Introduction}\label{sec:paper3_introduction}
When two massive main-sequence (MS) stars (\ie intermediate- and high-mass stars, with initial masses $M_{\mathrm{i}}$ of $1.3\,\msun \lesssim M_{\mathrm{i}} \lesssim 8\,\msun$ and $M_{\mathrm{i}} \gtrsim 8\,\msun$, respectively) merge, they form a new MS star with potentially peculiar properties. For example, it has been proposed and shown that such mergers produce strong, large-scale surface magnetic fields in the resulting merger products \citep{Ferrario2009,Wickramasinghe2014,Schneider2019, Ryu2025}. Should it indeed be true that such merger products are slow rotators, as found in \citet{Schneider2019} and \citet{Schneider2020}, they are a natural explanation for the blue MS band in young stellar clusters \citep{Wang2022}. MS merger products can also appear as blue stragglers in star clusters \citep{Rasio1995a,Sills1997,Sills2001,Mapelli2006,Glebbeek2008a,Ferraro2012,Schneider2015}. Further along their evolution, MS merger products can appear as red stragglers in a population of red supergiants, which can lead to cluster age underestimations of ${\sim}60\%$ \citep{Britavskiy2019}.

Despite these peculiarities, it is currently not straightforward to distinguish massive MS merger products from genuine single MS stars based on surface diagnostics alone. If one or more unambiguous distinguishing features of merger products were to be found, they could be used to confirm, for example, their slow-rotation hypothesis. To find such distinguishing features, we ought to go beyond the stars' surface diagnostics and assess any differences in their internal structure as predicted by merger simulations \citep{Lombardi1996,Sandquist1997,Sills2001,Freitag2005,Dale2006,Glebbeek2013,Schneider2019,Ballone2023}. Asteroseismology has proven to be the ideal tool to do so (see, \eg \citealt{Hekker2017, Aerts2021review, Kurtz2022, Bowman2023} for recent reviews). \citet{Bellinger2024} and \citet{Henneco2024b} made asteroseismic predictions to identify distinguishing features of different physical object classes appearing as blue supergiants, including post-MS merger products. \citet{Wagg2024} used asteroseismic predictions of rejuvenated MS accretors to assess whether they can be distinguished from MS stars with the same mass that have not accreted matter. They conclude that the effects of accretion on the accretor's internal structure produce a measurable difference in its asteroseismic signal compared to that of regular MS stars.

An obvious prerequisite for using asteroseismology is that the stars show pulsations. This is indeed the case for many massive MS stars. Thanks to space-based asteroseismic observations with, for example, Convection, Rotation and planetary Transits \citep[CoRoT,][]{Auvergne2009}, \textit{Kepler}/K2 \citep{Koch2010}, and the Transiting Exoplanet Survey Satellite \citep[TESS,][]{Ricker2016} a wealth of MS pulsators have been found and characterised \citep{Aerts2021review, Kurtz2022}. Fewer detections are currently available for stars with masses above roughly $8\,\msun$, which is the mass regime this work focuses on. Yet, the future looks bright thanks to the ongoing TESS and upcoming PLAnetary Transits and Oscillations of stars \citep[PLATO,][]{Rauer2024} missions. Stars in this mass regime have also been shown to exhibit low-frequency stochastic variability, for which the origin is currently still being debated \citep[\eg][]{Bowman2019, Bowman2024, Anders2023, Ma2024}. Mode excitation calculations \citep[\eg][]{Bouabid2013, Moravveji2016, Szewczuk2017} also predict massive MS stars to exhibit a variety of pulsations. Moreover, it is becoming increasingly clear that our current excitation theories tend to under-predict the number of excited linear modes \citep[\eg][]{Moravveji2016,Rehm2024}, as well as the actually observed modes in stars \citep[\eg][]{Balona2024,Hey2024}. Additional mode excitation theories for MS stars, such as non-linear resonant mode coupling \citep{Guo2022,VanBeeck2024}, are currently not included in mode instability predictions while such modes were found to be common among B-type pulsators \citep{VanBeeck2021}. Finally, tidal excitation \citep[][for a review]{Guo2021}
should not be ignored given the high fraction of massive stars in close binaries \citep{Sana2012}.  

This work consists of two parts. In the first part, we determine whether it is possible to distinguish a massive MS merger product from a genuine single MS star following a quasi-identical evolution in the Hertzsprung-Russell diagram (HRD). To do so, we make use of the 3D magnetohydrodynamic (MHD) merger product model from \citet{Schneider2019}, which we map to 1D following \citet{Schneider2020} to follow its post-merger evolution. In the second part, we repeat the first part's analysis, using 1D merger prescriptions instead of a 3D merger product model. 3D simulations of stellar mergers are computationally expensive, and hence, a limited number of 3D merger models are available. Multiple 1D merger prescriptions have been developed in an attempt to alleviate this problem. Contrary to 3D simulations, these 1D merger prescriptions do not model the merger phase itself but instead predict the structure of the merger product based on those of the binary components before the merger. Therefore, in the second part of this work, we investigate whether using three of these 1D merger prescriptions (entropy sorting, \texttt{Python Make Me A Massive Star}, and fast accretion) results in similarly structured merger product models as the one resulting from the 3D simulation. We then assess to what extent any asteroseismic differences between the MS merger product and its corresponding genuine single star found in the first part of this work are recovered with the 1D merger models.

The structure of this work is as follows. In Sect.~\ref{sec:paper3_diagnostics}, we provide the basic concepts and diagnostics of asteroseismology necessary for the analysis and discussion. Section \ref{sec:paper3_methods} covers the methods used to create merger products, evolve them and their genuine single-star counterparts, and predict their pulsations. We show and discuss our results in Sect.~\ref{sec:paper3_results}, and the conclusions can be found in Sect.~\ref{sec:paper3_dandc}. 

\section{Asteroseismic diagnostics}\label{sec:paper3_diagnostics}
This section gives a brief overview of some essential concepts and diagnostics of asteroseismology. We use these diagnostics to compare the asteroseismic predictions of a merger product and a genuine single star in Sect.~\ref{sec:paper3_results}.

The behaviour of pulsation modes depends on their dominant restoring force. Pressure (p) modes have the pressure gradient as their restoring force, while gravity (g) modes have buoyancy as their main restoring force. In rotating stars, the Coriolis force can also act as a restoring force alone (inertial waves) or in unison with the buoyancy force (gravito-inertial waves or GIWs). In slowly- and non-rotating stars, we describe the 3D geometry of pulsation modes with spherical harmonics $Y_{\ell}^{m}$ \citep{Aerts2010book}. The spherical degree $\ell$ ($\ell \geq 0$) gives the number of nodal lines (lines where the wave displacement is zero) on the stellar surface. Modes with $\ell > 0$ are called non-radial modes and are the main focus of this work. The azimuthal order $m$ ($|m| \leq \ell$) indicates how many of these nodal lines cross the equator. The radial order or overtone $n$ gives the number of nodal surfaces of a mode inside the star. We indicate the radial order of g and p modes with $n_{\mathrm{g}}$ and $n_{\mathrm{p}}$, respectively.

The p and g modes can only propagate in specific regions within the star, referred to as mode cavities. Outside of these cavities, in the evanescent zones, the modes decay exponentially. These p- and g-mode cavities are determined by the (linear\footnote{We use a tilde to emphasise that we use the linear definitions of the Lamb and BV frequencies instead of their often used angular forms. They are related to each other as $N = 2\pi \Tilde{N}$ and $S_{\ell} = 2\pi \Tilde{S}_{\ell}$.}) Lamb frequency $\Tilde{S}_{\ell}$ and (linear) Brunt-V\"{a}is\"{a}l\"{a} (BV) or buoyancy frequency $\Tilde{N}$, respectively. The BV frequency is defined as \citep{Aerts2010book}
\begin{equation}\label{eq:paper3fullbrunt}
    \Tilde{N}^2 = \frac{g}{4\pi^{2}}\left(\frac{1}{\Gamma_{1,\,0}}\frac{\mathrm{d}\ln P}{\mathrm{d}r} - \frac{\mathrm{d}\ln \rho}{\mathrm{d}r}\right)\,,
\end{equation}
or in approximate form for a fully ionized gas
\begin{equation}\label{eq:paper3approximatebrunt}
    \Tilde{N}^2 \simeq \frac{g^2\rho}{4\pi^{2}P}\left(\nabla_{\mathrm{ad}} - \nabla + \nabla_{\mu} \right)\,.
\end{equation}
In these expressions, $g$ is the local gravitational acceleration, $\Gamma_{1,\,0}$ is the first adiabatic exponent, $P$ is the pressure, $\rho$ is the density, $r$ is the radial coordinate, and
\begin{equation}
    \nabla = \frac{\mathrm{d}\ln T}{\mathrm{d}\ln P}\,,\quad \nabla_{\mathrm{ad}} = \left(\frac{\mathrm{d}\ln T}{\mathrm{d}\ln P}\right)_{\mathrm{ad}}\,,\quad \nabla_{\mu} = \frac{\mathrm{d}\ln \mu}{\mathrm{d}\ln P}\,,
\end{equation}
with $T$ the temperature, $\mu$ the mean molecular weight, and the `ad' subscript referring to the adiabatic assumption. The Lamb frequency is defined as \citep{Aerts2010book}

\begin{equation}\label{paper3Lamb}
    \Tilde{S}_{\ell}^2 = \frac{\ell(\ell+1) c_{\mathrm{s}}^2}{4\pi^{2}r^2}\,,
\end{equation}
with $c_{\mathrm{s}}$ the local sound speed. 

The g-mode cavity is determined by $|\nu| < |\Tilde{N}|$ and $|\nu| < \Tilde{S}_{\ell}$, with $\nu$ the linear mode frequency, while p modes with frequency $\nu$ can only propagate when $|\nu| > |\Tilde{N}|$ and $|\nu| > \Tilde{S}_{\ell}$. 

In the asymptotic regime, that is, for $n \gg 1$, g modes of the same $\ell$ and consecutive radial orders $n_{\mathrm{g}}$ are equally spaced in period when the star is non-rotating, non-magnetic, and chemically homogeneous \citep{Tassoul1980}. The spacing between the periods of high-order g modes, $\Delta P_{n} = P_{n} - P_{n-1}$, with $P_{n}$ the mode period of a mode with radial order $n$, is then equal to the asymptotic period spacing $\Pi_{\ell}$, defined as \citep{Aerts2010book}
\begin{equation}\label{eq:paper3Pil}
    \Pi_{\ell} = \frac{\Pi_{0}}{\sqrt{\ell(\ell+1)}}\,,
\end{equation}
with
\begin{equation}\label{eq:buoyancy_travel_time}
    \Pi_{0} = \pi\left(\int_{r_{\mathrm{i}}}^{r_{\mathrm{o}}}\frac{\Tilde{N}}{r}\mathrm{d}r\right)^{-1}
\end{equation}
the characteristic period, also termed the buoyancy travel time. In this expression, $r_{\mathrm{i}}$ and $r_{\mathrm{o}}$ are the radial coordinates at the inner and outer boundaries of the g-mode cavity, respectively. If a so-called period spacing pattern (PSP), that is, $\Delta P_{n}$ as a function of $n_{\mathrm{g}}$ or $P_{n}$, is observable, it can be used to estimate the size of the g-mode cavity and hence the size of the convective core \citep{Moravveji2015,Moravveji_etal2016,Pedersen2018,Pedersen2021,Mombarg2019,Mombarg2021}. Additionally, since many stars are not chemically homogeneous, rotate, and have magnetic fields, their g modes are not equally spaced. Departures of the PSP from the constant value $\Pi_{\ell}$ hold a tremendous amount of information about the stellar interior. For example, structural and chemical glitches, which influence $\Tilde{N}$, can lead to quasi-periodic variation in $\Delta P_{n}$ \citep{Miglio2008,Degroote2010,Cunha2015,Cunha2019,Cunha2024}. The Coriolis force in a rotating star will introduce a slope in the PSP depending on $\ell$, $m$, and the angular rotation frequency $\Omega$ in the mode cavity \citep{Bouabid2013}, as observed in hundreds of MS pulsators meanwhile \citep[\eg][]{VanReeth2015a, VanReeth2015b, VanReeth2016, Li2020, Pedersen2021, Szewczuk2021, Garcia2022b}.

\section{Methods}\label{sec:paper3_methods}

\subsection{Stellar evolution computations with \texttt{MESA}}\label{sec:mesa}
We used the 1D stellar structure and evolution code \texttt{MESA} \citep[r12778,][]{Paxton2011, Paxton2013, Paxton2015, Paxton2018, Paxton2019} to compute the input models for the various merger prescriptions and evolve the resulting merger products. We computed the genuine single-star models using the same \texttt{MESA} setup and, hence, input physics. The choices for the input physics and setup were based on those from \citet{Schneider2020}, except that we did not include rotation at the level of the equilibrium models, we did not model any accretion from the disk formed during the merger event, and we ignored the magnetic field produced in the merger process \citep{Schneider2019}. The number of works on the effect of magnetic fields on non-radial pulsations has been growing steadily (see, \eg \citealt{Prat2019}, \citealt{VanBeeck2020}, \citealt{Dhouib2022}, \citealt{Rui2024}, \citealt{Bessila2024}, \citealt{Bhattacharya2024}, and \citealt{Hatt2024}). However, we aim to assess the effects of the structure and composition of MS merger products separately from the effects of magnetic fields. Ignoring rotation in the equilibrium models and only taking it into account at the level of the oscillation equations (see Sect.~\ref{sec:gyre}) is justified by the small effect of the centrifugal deformation of the star on predicted frequencies \citep{Henneco2021,Dhouib2021} and is common practice for slow to moderate rotators \citep{Aerts2021review,AertsTkachenko2024}. We compensated for the resulting lack of rotationally induced mixing by mimicking its effect with a constant envelope mixing of $\log (D_{\mathrm{mix}}/\mathrm{cm}^{2}\mathrm{s}^{-1}) = 3$. This envelope mixing was also used to smooth out small chemical glitches introduced during the merger and left behind by the receding convective core during the MS evolution. This is a typical value for envelope mixing inferred from asteroseismic modelling of single B-type stars \citep{Pedersen2021,Burssens2023}. 

We used mixing length theory \citep[MLT,][]{Bohm-vitense1958,Cox1968} to treat convection in our models with a mixing length parameter of $\alpha_{\mathrm{mlt}} = 1.8$. We assessed the stability against convection using the Ledoux criterion. Additional mixing was included in the form of thermohaline mixing with an efficiency of $\alpha_{\mathrm{th}} = 2.0$ and semi-convective mixing with an efficiency of $\alpha_{\mathrm{sc}} = 1.0$ (semi-convective mixing only appears in our models during thermal relaxation phases, not during the MS evolution). We used the exponential overshoot scheme to account for convective boundary mixing above convective cores with $f_{\mathrm{ov}} = 0.019$. This value for $f_{\mathrm{ov}}$ was based on observational constrains from ${\approx}10\,\msun$ MS stars by \citet{Castro2014} \citep[see also][]{Schneider2020}. We used the \citet{Vink2001} wind mass-loss prescription with a scaling factor of one (we only consider the MS evolution, \ie the hot star regime for wind mass loss). All models were computed at solar metallicity ($Y = 0.2703$ and $Z=0.0142$, \citealt{Asplund2009}), with a combination of the OPAL \citep{Iglesias1993,Iglesias1996} and \citet{Ferguson2005} opacity tables suitable for the chemical mixture from \citet{Asplund2009}. The models were terminated when the central hydrogen mass fraction $X_{\mathrm{c}}$ dropped below $10^{-6}$.

In this work, we conduct a comparative analysis between models using the same exact set of input physics and assumptions. Therefore, our choices of uncertain physical processes, such as convective boundary mixing, would only matter if they are different in genuine single-star and merger product models. Since we assume they are the same in both types of objects, changing any of these parameters would affect both the single-star and merger product models but leave the systematic differences intact.

\subsection{Merger models and prescriptions}\label{sec:mergermethods}
This section describes how we obtained a 1D model for an MS merger product from the 3D MHD simulation from \citet{Schneider2019} and three 1D merger prescriptions used to approximate this model. The 3D MHD simulation, as well as the three 1D methods, used $9\,\msun$ and $8\,\msun$ single-star 1D \texttt{MESA} models evolved up to $9\,$Myr. At this point, their central hydrogen mass fractions were $X_{\mathrm{c}} = 0.60$ and $X_{\mathrm{c}} = 0.62$, respectively, and the mass ratio $q=M_{2}/M_{1} = 0.89$. A limitation that all of the methods described below share, including the 3D MHD merger product model, is that mass transfer before and during the contact phase, which precedes the stellar merger, is not included in this initial study of MS merger asteroseismology. From detailed binary evolution calculations, such as those from \citet{Pols1994}, \citet{Wellstein2001}, \citet{deMink2007}, \citet{Claeys2011}, \citet{Marchant2016}, \citet{Mennekens2017}, \citet{Laplace2021}, \citet{Menon2021}, and \citet{Henneco2024a}, we know that mass transfer can significantly alter the structure of the binary components. The impact of ignoring the mass-transfer phase prior to merging will be assessed in a future study (Heller et al. in prep.).

\subsubsection{3D MHD model}\label{sec:methods:mhd}
Following \citet{Schneider2019} and \citet{Schneider2020}, we started from the chemical composition and entropy profiles of the $16.9\,\msun$ merger product resulting from the 3D MHD simulation of \citet{Schneider2019} computed with the moving-mesh code \textsc{Arepo} \citep{springel2010a, pakmor2011a}. These were used to relax a 1D stellar model in MESA with the same total mass, chemical composition profile, and entropy structure (\ie thermal structure) as the 3D merger product (see Appendix B of \citealt{Paxton2018} for a technical description of this relaxation routine). The resulting 1D model was then used as the starting point of a \texttt{MESA} evolution run with the physical and numerical choices described in Sect.~\ref{sec:mesa}. As described in \citet{Schneider2019,Schneider2020}, the $8\,\msun$ secondary star's core sinks to the centre of the merger product, and the $9\,\msun$ primary star's more evolved and more He-rich core ends up in the layer around it. Consequently, the merger product's inner envelope is enriched in helium (He) and other products of hydrogen (H) burning. During the initial phases of the evolution of the merger product, it thermally relaxes, leading to a rapid expansion and subsequent contraction phase. During this contraction phase, the merger product has a transient ($\Delta t \simeq 9000\,$yr) convective core reaching a mass of ${\approx}11\,\msun$ (for comparison, when the star arrives again on the MS after thermally relaxing, its convective core mass ${\approx}7\,\msun$). As detailed in \citet{Schneider2020}, the carbon (C) and nitrogen (N) abundances in the core are out of equilibrium, and the core is denser and hotter than in full equilibrium after the merger. These two conditions lead to a phase of enhanced nuclear burning and are, therefore, the origins of the transient convective core. We refer to this merger product model as the `3D MHD merger product' or the `3D MHD model' to distinguish it from the 1D merger product models. The corresponding acronym in figures, sub-, and superscripts is `MHD'. This state-of-the-art simulation is probably the most accurate structure of a merged star after the dynamic coalescence currently available. It thus serves as a benchmark model in this paper. The biggest uncertainty on the resulting stellar structure stems from the mapping back into MESA \citep[see][]{Schneider2019}. The 3D merger structure consists of a $3\,\msun$ rotationally-supported torus around the spherically-symmetric central merger remnant and its evolution through accretion of the torus material into the final merged star remains uncertain. This mostly affects the outermost layers of the merged star and the star's rotational profile.

\subsubsection{Entropy sorting}
Entropy sorting is based on the relation between the entropy and buoyancy of stellar material. In an ideal case, a star in hydrostatic equilibrium has a monotonically increasing entropy profile, except for convective regions, where the entropy profile is approximately flat \citep{Lombardi1996}. The layers with lower entropy have lower buoyancy and are thus found closer to the star's centre. In a simplified picture, when two stars merge, the layers with the lowest entropy will sink to the centre of the merger product. Using this principle, we combined the structures of the $9\,\msun$ and $8\,\msun$ progenitor stars based on their entropy profiles. Starting from the centre, we selected the layer from either star with the lowest entropy, creating a new structure with a monotonically increasing entropy profile. Analogous to the 3D MHD model, we used the chemical composition profile to relax a \texttt{MESA} model with a total mass of $16.9\,\msun$. Although it might seem logical to use the entropy-sorted entropy profile as well for the relaxation routine, doing so leads to abnormally high central temperatures and densities in the relaxed model. This is caused by the fact that the entropy sorting model is not in hydrostatic equilibrium prior to relaxing, that is, the central entropy has not adjusted to the merger product's mass. For the 3D MHD merger product and \texttt{PyMMAMS} model this is not a problem since they are in hydrostatic equilibrium. Also, contrary to the 3D MHD model, the chemical composition profiles resulting from entropy sorting is that of a $17.0\,\msun$ stellar model. In other words, we did not yet account for the $0.1\,\msun$ lost during the merger in the 3D MHD simulation. Since we did want to make assumptions for the composition of the lost material during the merger, we relaxed the original $17.0\,\msun$ \texttt{MESA} model to a $16.9\,\msun$ model, as opposed to stripping the $0.1\,\msun$ from the merger product's surface after relaxation. The relaxed model was subsequently evolved in \texttt{MESA}. We did not employ any artificial smoothing of the merger product's structure; small chemical and structural glitches were smoothed out during the relaxation phase and subsequent MS evolution because of the envelope mixing described in Sect.~\ref{sec:mesa}. We refer to the merger product constructed through entropy sorting as the `entropy-sorted model' or `entropy-sorted merger product', with the acronym `ES'.

A limitation of the entropy sorting method is that it does not consider any form of entropy generation. During the merging phase, shocks can increase the entropy in both companions and, in general, additional mixing occurs. We see the consequences of this in our case study of the merger between the $9\,\msun$ and $8\,\msun$ stars. Assuming the stars are born together, the more massive $9\,\msun$ primary star has a more evolved, He-rich core with a lower mean entropy than the core of the $8\,\msun$ secondary star. By applying entropy sorting, we thus find that the core of the primary sinks to the centre of the merger product (this is further described in Sec.~\ref{sec:ES_results}), whereas we found from the 3D MHD model that the secondary's core has sunk to the centre.

\subsubsection{\texttt{PyMMAMS}: entropic variable sorting with shock heating}\label{sec:methods_pymmams}
The \texttt{Make Me A Massive Star} (\texttt{MMAMS}) routine is a 1D merger prescription originally presented in \cite{Gaburov2008b} and includes an approximation for the shock heating (entropy generation) that occurs during stellar head-on collisions (as opposed to slower inspiral mergers driven by binary evolution). \texttt{MMAMS} performs stellar mergers by first shock heating the progenitors and then sorting them using the entropic variable $A$ \citep{Gaburov2008b}, 
 \begin{equation}
 A = \frac{\beta_{\mathrm{gas}} P}{\rho^{5/3}} \exp\left[\frac{8}{3}\frac{\left(1-\beta_{\mathrm{gas}}\right)}{\beta_{\mathrm{gas}}}\right],
 \end{equation}
which is closely linked to the specific entropy. Here, $\beta_{\mathrm{gas}}$ is the ratio of gas pressure $P_{\mathrm{gas}}$ over total pressure $P$. The entropic variable is used together with a pressure estimate from hydrostatic equilibrium to compute the density of the progenitor shells in the merger product. \texttt{MMAMS} builds the merger remnant starting at the center, with the more dense shells being placed closer to the core. The merger product is then dynamically relaxed by solving the equations of hydrostatic equilibrium. Shock heating changes the entropic variable profile of the stars, leading to changes in the post-merger composition profiles compared to entropy sorting. The shock heating prescription was obtained from smoothed particle hydrodynamic (SPH) simulations of stellar head-on collisions for progenitors of different initial masses, mass ratios and evolutionary stages \citep{Gaburov2008b}. The prescription includes a correction factor to account for energy conservation before and after the merger. 

Currently, \texttt{MMAMS} is available as part of the \texttt{AMUSE} framework \citep{Portegieszwart2009,Pelupessy2013,Portegieszwart2013,Portegieszwart2018,Portegieszwart2019}. For better portability and modifiability, we translated \texttt{MMAMS} to \texttt{Python} (Heller et al., in prep.). It is this \texttt{Python} version, \texttt{Python Make Me A Massive Star} (\texttt{PyMMAMS}), that we used in this work. We implemented several modifications in \texttt{PyMMAMS} compared to the original \texttt{MMAMS} code. For example, we introduced a new re-meshing scheme, which alleviates the double-valuedness in regions of the merged star, where progenitor mass elements of very different compositions ended up next to each other. Our re-meshing scheme aims to improve upon the mixing scheme included in the original code, which mixed stellar matter over steep composition gradients. These gradients are located at the interface of single- and double-valued regions, that is, parts of the merger product where shells of material coming from only one of the progenitors touch shells of material consisting of a mixture of both progenitors. In the original mixing scheme from \texttt{MMAMS}, these gradients were softened, which in some cases led to hydrogen from the envelope being mixed into the helium core of a post-MS merger product, distorting the merger product's further evolution. 

Certain numerical solvers used to compute the shock heating in \texttt{(Py)MMAMS} have been found to fail for mass ratios $q=M_{2}/M_{1}$ below 0.1 and above 0.8. Since the mass ratio of our progenitor binary system is $q = 0.89$, we compute the shock heating for a $q=0.8$ and use that to alleviate this shortcoming of \texttt{(Py)MMAMS}.

Even though \texttt{PyMMAMS} includes the effect of shock heating, the shock-heating prescription has been calibrated for head-on collisions. These tend to be more energetic than the slower inspiral of binary components of a stellar merger driven by binary evolution. Therefore, we consider \texttt{PyMMAMS} and entropy sorting to be the two extremes regarding shock heating, with the actual amount of shock heating likely somewhere in between. 

We refer to the merger product model obtained with \texttt{PyMMAMS} as the `\texttt{PyMMAMS} model' or `\texttt{PyMMAMS} merger product'. The corresponding acronym is `PM'.

\subsubsection{Fast accretion}
The last 1D merger method used in this work is fast accretion. This method consists of accreting a certain amount of mass onto a star, in this case, the $9\,\msun$ primary star, in a timescale shorter than or equal to the primary star's global thermal timescale. We closely followed the setup of \citet{Henneco2024b} and accreted $7.9\,\msun$ of material with the same chemical composition as the surface of the primary star during 10\% of the primary star's global thermal timescale $\tau_{\mathrm{KH}}$. We initiated this fast accretion phase when the primary star reached an age of $9\,$Myr. The main limitations of this method are described in \citet{Henneco2024b}. We describe its restrictions specifically for reproducing massive MS merger products in Sects.~\ref{sec:FA_results} and \ref{paper3:improvements}. The merger product constructed with the fast accretion method is referred to as the `fast accretion model' or `fast accretion merger product'. For this model, we use the acronym `FA'.

\subsection{Oscillation mode predictions with \texttt{GYRE}}\label{sec:gyre}
We used the stellar oscillation code \texttt{GYRE} \citep[v7.0;][]{Townsend2013,Townsend2018} to predict the oscillation modes for the equilibrium models obtained from the 1D \texttt{MESA} models described above. We used the \texttt{MAGNUS\_GL6} solver with the boundary conditions from \citet{Unno1989book}. For g modes in the absence of rotation, we computed $(\ell,\,m) = (1,\,0)$ and $(\ell,\,m) = (2,\,0)$ modes. For predictions with rotation, we used the traditional approximation of rotation \citep[TAR,][]{Eckart1960,Berthomieu1978,Lee1987,Townsend2003,Mathis2009} as implemented in \texttt{GYRE} to compute $(\ell,\,m) = (1,\,0)$, $(\ell,\,m) = (1,\,\pm 1)$, $(\ell,\,m) = (2,\,0)$, $(\ell,\,m) = (2,\,\pm 1)$, and $(\ell,\,m) = (2,\,\pm 2)$ modes in the inertial (observer's) frame. All these computations of g modes were conducted with the adiabatic approximation, which is appropriate to compute the frequencies of g modes in B-type stars \citep{Aerts2018b}. 

We computed p modes under non-adiabatic conditions because they are more sensitive to the star's outer layers, where the thermal timescale becomes relatively short and non-adiabatic effects may become important. We used the \texttt{MAGNUS\_GL2} solver together with \texttt{GYRE}'s \texttt{CONTOUR} initial search method for these non-adiabatic p-mode computations. As stated by the \texttt{GYRE} documentation, the \texttt{MAGNUS\_GL2} solver is more appropriate for non-adiabatic computations, given that it does not lead to convergence issues. In the absence of rotation, we computed p modes of $(\ell,\,m) = (1,\,0)$ and $(\ell,\,m) = (2,\,0)$. The effects of rotation were included through the first-order Ledoux perturbative approach (see \citealt{AertsTkachenko2024} for more details on this approach) for $(\ell,\,m) = (1,\,\pm 1)$, $(\ell,\,m) = (2,\,\pm 1)$, and $(\ell,\,m) = (2,\,\pm 2)$ p modes. We did not add atmosphere models to the equilibrium model output used by \texttt{GYRE} because the MS stars treated in this work are not expected to have extended atmospheres.

\section{Results and discussion}\label{sec:paper3_results}

\begin{figure}
    \centering
    \resizebox{0.90\hsize}{!}{\includegraphics{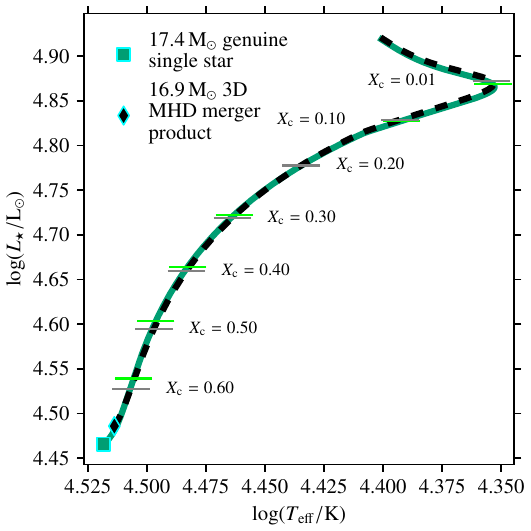}}
    \caption{HRD with MS evolutionary tracks for the $16.9\,\msun$ 3D MHD merger product (black dashed line) and the $17.4\,\msun$ genuine single star (green solid line). The grey and lime-coloured horizontal markers indicate different evolutionary stages, using the central hydrogen mass fraction $X_{\mathrm{c}}$ as a proxy for the evolutionary age, for the merger product and genuine single star, respectively.}
    \label{fig:hrd_mhd}
\end{figure}

\begin{figure*}
\centering
  \includegraphics[width=15cm]{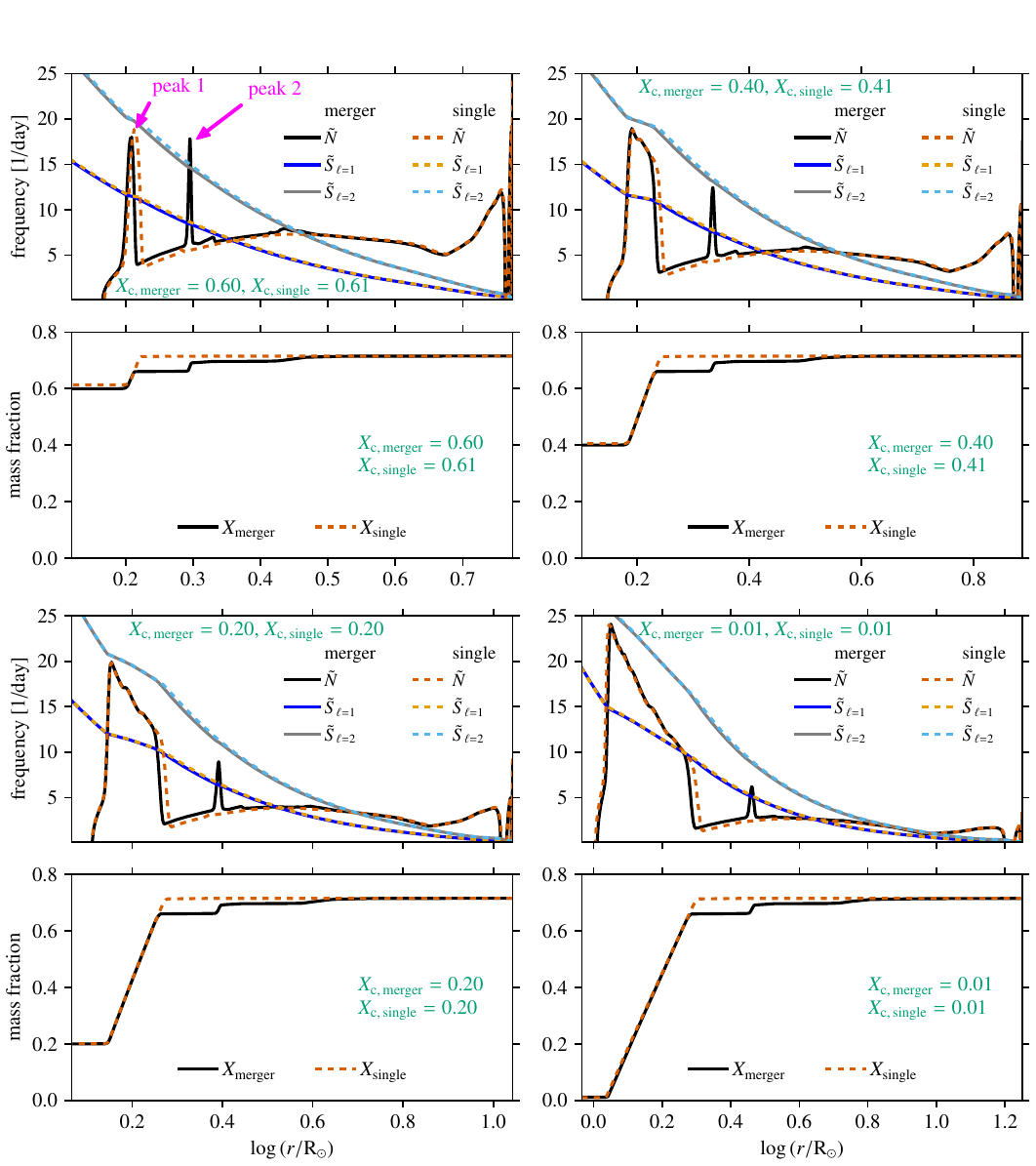}
     \caption{Evolution of the propagation diagrams (odd rows) and hydrogen mass fraction X profiles (even rows) for the $16.9\,\msun$ 3D MHD merger product (solid lines) and the $17.4\,\msun$ genuine single star (dashed lines). Each panel is labelled with the corresponding central hydrogen mass fractions of the merger product and the genuine single star.}
     \label{fig:mhd_props_and_comps_evolution}
\end{figure*}

Figure \ref{fig:hrd_mhd} shows the MS evolutionary tracks of the $16.9\,\msun$ merger product constructed from the 3D MHD simulation described in Sect.~\ref{sec:methods:mhd}, and a genuine single $17.4\,\msun$ star in an HRD. We chose a mass of $17.4\,\msun$ for the genuine single-star model because we found that its evolutionary track in the HRD overlaps almost completely with that of the $16.9\,\msun$ merger product. Since the massive MS merger product's HRD track overlaps with a higher-mass genuine single-star track ($\Delta M_{\star} = 0.5\,\msun$), we find that the merger product has a higher $L_{\star}/M_{\star}$ compared to the genuine single star ($L_{\star}$ and $M_{\star}$ are the luminosity and total mass of the star, respectively). This higher $L_{\star}/M_{\star}$ follows from the fact that the He-rich core material of the more evolved star ends up in the lower envelope of the merger product, leading to a higher mean molecular weight $\mu$ there. This can be inferred from the chemical composition profiles in Fig.~\ref{fig:mhd_props_and_comps_evolution}. Following the mass-luminosity relation for MS stars, $L_{\star} \propto M_{\star}^{3}\mu^{4}$ \citep{Kippenhahn2012}, we see that the He enrichment compensates for the merger product's lower mass. In addition to the similar values in luminosity and effective temperature, we see from Fig.~\ref{fig:LRMcc_mhd} that throughout their MS evolution, the merger product and genuine single star also have similar radii $R_{\star}$ and convective core radii $R_{\mathrm{cc}}$ when they occupy the same position in the HRD. We emphasise that this behaviour in $L_{\star}/M_{\star}$ is generic, can be brought back to the basic mass-luminosity and mass-radius relations, and should be expected for any star with chemically enhanced envelopes. 

From the horizontal bar markers on the HRD tracks in Fig.~\ref{fig:hrd_mhd}, we see that for the same effective temperature $T_{\mathrm{eff}}$ and luminosity $L_{\star}$, the two stars are at different MS ages, that is, they have different central hydrogen mass fractions $X_{\mathrm{c}}$. Initially, the genuine single star still has more hydrogen in its convective core than the merger product when they are in the same position in the HRD. With increasing time, this difference in $X_{\mathrm{c}}$ becomes smaller, reaching a minimum around the time when both stars have $X_{\mathrm{c}}\approx 0.20$. At later times, the merger product has a higher value for $X_{\mathrm{c}}$ than the genuine single star for the same $T_{\mathrm{eff}}$ and $L_{\star}$. 

\begin{figure}
    \centering
    \resizebox{0.85\hsize}{!}{\includegraphics{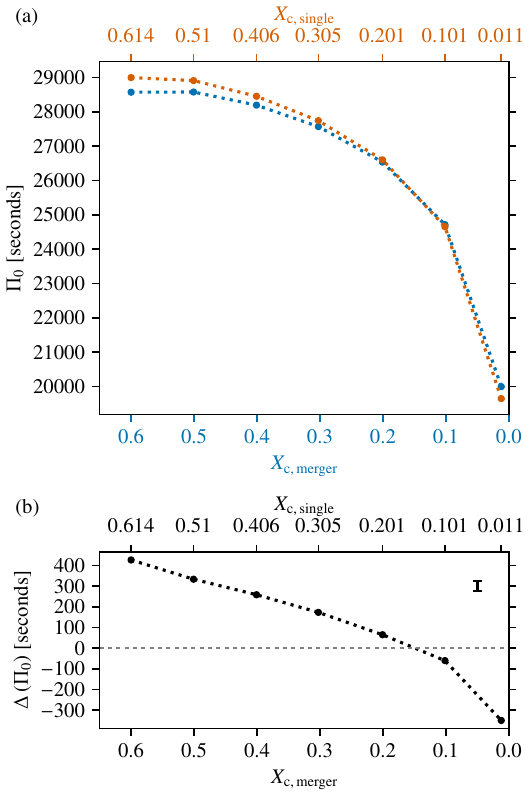}}
    \caption{Comparison between the buoyancy travel time $\Pi_{0}$ of the $16.9\,\msun$ 3D MHD merger product (blue line) and the $17.4\,\msun$ genuine single star (red line) as a function of their respective central hydrogen mass fractions $X_{\mathrm{c}}$ (Panel a). Panel (b) shows the absolute differences in $\Pi_{0}$, $\Delta (\Pi_{0}) = \Pi_{0,\,\mathrm{single}} - \Pi_{0,\,\mathrm{merger}}$. The error bar in panel (b) shows $\sigma_{\Delta P}$.}
    \label{fig:Pi_vs_Xc_matched}
\end{figure}

Although it is instructive to compare the asteroseismic predictions for our models at certain values of $X_{\mathrm{c}}$ (as done, for example, in \citealt{Wagg2024}), we opted to compare models when they have similar luminosities and effective temperatures. We make this choice because we want to identify distinguishing features in the asteroseismic predictions for a merger product and genuine single star with similar surface diagnostics, which we get from observations. Therefore, in practice, we compare the models at specific $X_{\mathrm{c}}$ values for the merger product. At each of these points in the HRD, we compare the predictions for the merger product with those for the genuine single-star model closest in terms of $L_{\star}$ and $T_{\mathrm{eff}}$. In other words, we compare the models at the positions in the HRD marked by the grey horizontal markers in Fig.~\ref{fig:hrd_mhd}. The corresponding genuine single-star models have only slightly different values of $X_{\mathrm{c}}$ (at most $2.3\%$), which can be seen by comparing the top and bottom $x$-axes in Figs.~\ref{fig:LRMcc_mhd} and \ref{fig:Pi_vs_Xc_matched}.

\subsection{Asteroseismic comparison}\label{sec:results_astero_compare}
The $16.9\,\msun$ 3D MHD merger product and the $17.4\,\msun$ genuine single star find themselves in the region of the HRD associated with $\beta$~Cephei ($\beta$~Cep) pulsators \citep{Aerts2010book,Aerts2021review}. Stars in this class pulsate in numerous low-order p and g modes \citep[\eg][]{Burssens2023, Fritzewski2024b}, as well as in some high-order g modes \citep[\eg][]{Daszynska2017} when observed in modern space photometry. In this section, we compare the predictions for the g and p modes in this merger product and genuine single star, with and without rotation.

\begin{figure*}
\centering
  \includegraphics[width=16cm]{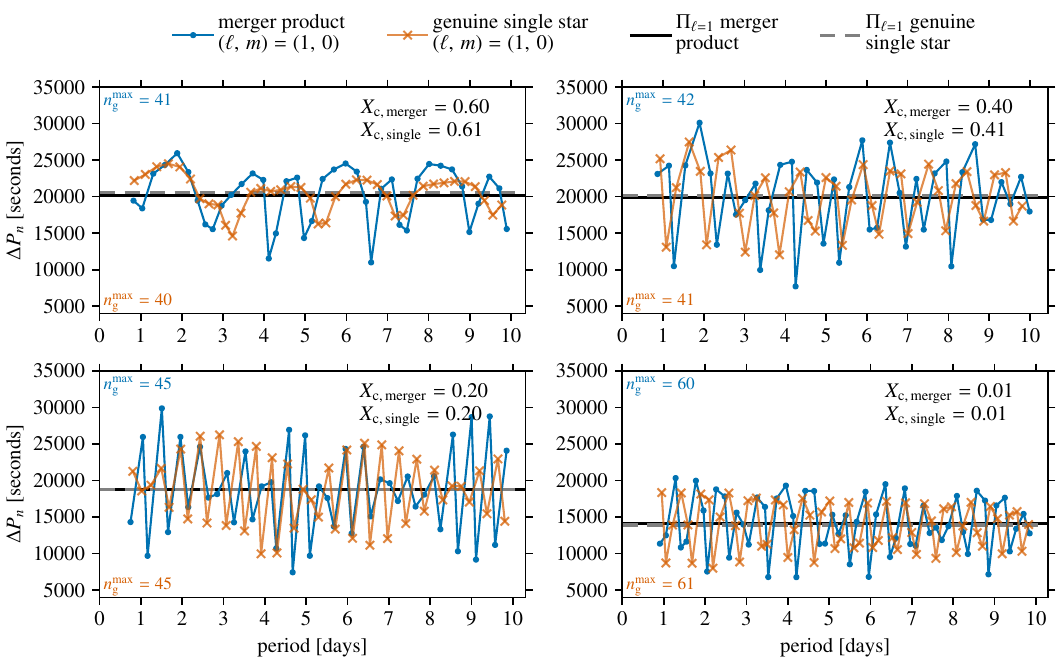}
     \caption{Period spacing patterns for $\ell = 1$ g modes without rotation for the $16.9\,\msun$ 3D MHD merger product (blue solid lines, dot markers) and the $17.4\,\msun$ genuine single star (orange solid lines, cross markers) at different evolutionary stages. The horizontal solid black and dashed grey lines indicate the values of the asymptotic period spacing values $\Pi_{\ell=1}$ for the merger product and genuine single star, respectively. The $n_{\mathrm{g}}^{\mathrm{max}}$ values show the maximum radial order plotted for the merger product (blue) and genuine single star (red).}
     \label{fig:psp_evolution}
\end{figure*}

\subsubsection{Gravity modes}\label{sec:paper3_gmodes}

\begin{figure*}[]
\centering
  \includegraphics[width=16cm]{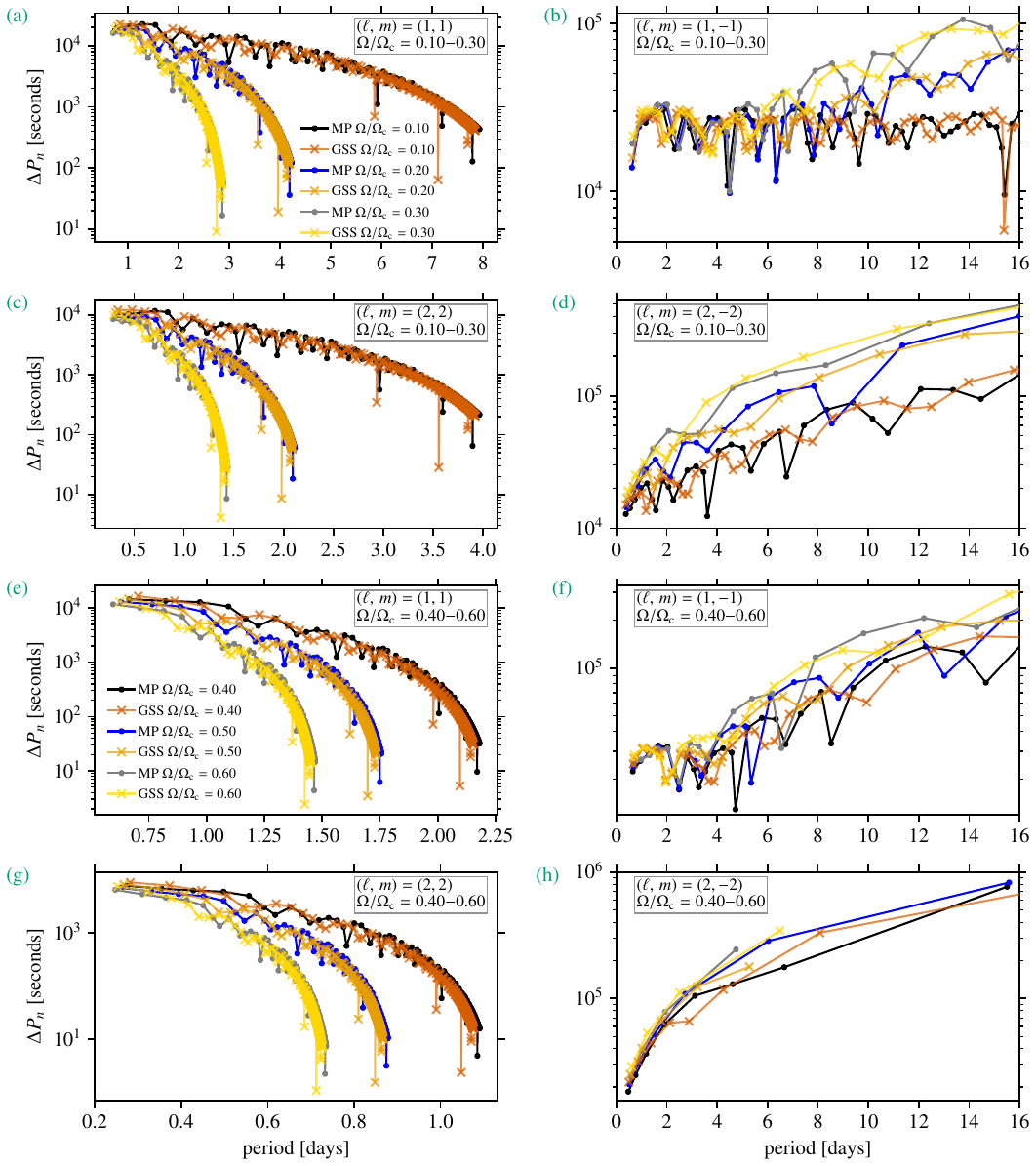}
     \caption{Period spacing patterns (PSPs) for $(\ell,\,m) = (1,\,\pm 1)$ and $(\ell,\,m) = (2,\,\pm 2)$ g modes with rotation rates of $\Omega/\Omega_{\mathrm{c}} = 0.10\text{--}0.30$ (Panels a--d) and $\Omega/\Omega_{\mathrm{c}} = 0.40\text{--}0.60$ (Panels e--h) for the $16.9\,\msun$ 3D MHD merger product and the $17.4\,\msun$ genuine single star at $X_{\mathrm{c,\,merger}} = 0.50$ and $X_{\mathrm{c,\,single}} = 0.51$, respectively. The black, blue, and grey lines with dot markers correspond to the merger product's PSPs, while the red, orange, and gold lines with cross markers correspond to the PSPs of the genuine single star. The PSPs are shown in the inertial (observer's) frame.}
     \label{fig:psp_rotation}
\end{figure*}

\begin{figure}
    \centering
    \resizebox{0.9\hsize}{!}{\includegraphics{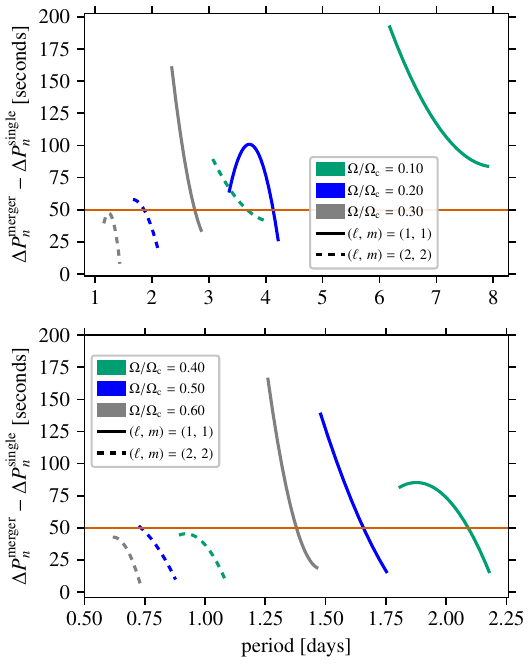}}
     \caption{Differences between the quadratic function fits of the PSPs for prograde modes shown in Fig.~\ref{fig:psp_rotation} for $\Omega/\Omega_{\mathrm{c}} = 0.10\text{--}0.30$ (top) and $\Omega/\Omega_{\mathrm{c}} = 0.40\text{--}0.60$ (bottom). The solid and dashed lines show the differences for the $(\ell,\,m) = (1,\,1)$ and $(\ell,\,m) = (2,\,2)$ modes, respectively. The horizontal red line shows the average value of $\sigma_{\Delta P}$ from observations of SPB pulsators.}
     \label{fig:psp_rotation_diffs}
\end{figure}

We start by looking at the difference between the buoyancy travel times $\Pi_{0}$ (Eq.~\ref{eq:buoyancy_travel_time}) for the two objects at different points along their evolution, shown in Fig.~\ref{fig:Pi_vs_Xc_matched}. We note that we integrate over all regions of the star where $\Tilde{N}^{2} > 0$ when computing $\Pi_{0}$. In other words, we integrate over both the main g-mode cavity and the sub-surface g-mode cavity caused by the iron opacity bump. \citet{Serebriakova2024} find that g modes are able to tunnel through the sub-surface convection zone separating these two mode cavities in stars with masses between $12$ and $30\,\msun$. Hence, integrating over both mode cavities is appropriate. The buoyancy travel times span a range of roughly $20\times 10^3\,$s to $30\times 10^3\,$s ($5.56\text{--}8.33\,$h). The absolute differences between the buoyancy travel times of the merger product and genuine single star (Fig.~\ref{fig:Pi_vs_Xc_matched}b) have a median value of $173\,$s and lie in the interval $[-349;\,426]\,$s. If we convert this to the asymptotic period spacing via Eq.~(\ref{eq:paper3Pil}), we find that the median values of the difference in $\Pi_{\ell}$ are $122\,$s and $71\,$s for $\ell=1$ and $\ell=2$, respectively. These values are considerably higher than the currently best uncertainties for observed period spacing values of B-type stars $\sigma_{\Delta P} = 50\,$s \citep{Degroote2010,Moravveji2015,Pedersen2021}. At earlier evolutionary stages (higher values of $X_{\mathrm{c}}$), the genuine single star has a longer buoyancy travel time than the merger product. Around $X_{\mathrm{c,\,merger}} \approx X_{\mathrm{c,\,single}} \approx 0.15$, the merger product overtakes the genuine single star in $\Pi_{0}$. In summary, we see that the $\Pi_{0}$ values differ for the merger product and genuine single star during large parts of their MS lifetimes, and this difference follows a trend with $X_{\mathrm{c}}$.\\

To explain these differences in $\Pi_{0}$ for these two types of stars, we compare their g-mode cavities shown in the propagation diagrams in Fig.~\ref{fig:mhd_props_and_comps_evolution}. We identify multiple differences. First, there is the location of the inner boundary of the mode cavity $r_{\mathrm{i}}$, which is equal to the convective core radius $R_{\mathrm{cc}}$. Since we integrate $\Tilde{N}/r$ to compute $\Pi_{0}$, the latter's value is the most sensitive to that of $\Tilde{N}$ at this inner boundary. From Fig.~\ref{fig:LRMcc_mhd}, we see that the absolute difference between $R_{\mathrm{cc,\,merger}}$ and $R_{\mathrm{cc,\,single}}$ is at most $0.0125\,\rsun$ and it follows a similar trend as $\Delta(\Pi_{0})$ in Fig.~\ref{fig:Pi_vs_Xc_matched}b. This follows from the fact that a larger convective core leads to a less extended (in radius) g-mode cavity. Second, the leftmost BV frequency peak (`peak 1' in the top left panel of Fig.~\ref{fig:mhd_props_and_comps_evolution}), caused by the chemical gradient left behind by the receding convective core, are wider for the genuine single star at all points along the MS. If this were the only difference between the two g-mode cavities, we would expect lower $\Pi_{0}$ values of the genuine single star than for the merger product. Third, to the right of this BV frequency peak (peak 1), we see that the genuine single star's $\Tilde{N}$ is lower across multiple solar radii, and the merger product has an additional peak (`peak 2') in its $\Tilde{N}-$profile. This peak results from the transient convective core of the merger product during its thermal relaxation phase described in Sect.~\ref{sec:methods:mhd}. The effect of $\Tilde{N}_{\mathrm{merger}} > \Tilde{N}_{\mathrm{single}}$ in this region is to lower $\Pi_{0}$ for the merger product compared to the one for the genuine single star. Considering all three effects together, we see that the differences in the merger product's and genuine single star's respective $\Pi_{0}$ follow the same trend as the convective core radius, but that the point at which the merger product's $\Pi_{0}$ becomes larger than the one of the single star occurs later in the evolution than for the convective core radius because of differences in $\Tilde{N}$ in the near-core and envelope regions of both models.\\

We now turn our attention to the PSPs for $\ell=1$ modes (those for $\ell=2$ modes are shown in Fig.~\ref{fig:psp_evolution_l2} in Appendix~\ref{app:l2modes}) in the absence of rotation. Figure \ref{fig:psp_evolution} shows PSPs constructed using \texttt{GYRE} predictions (see Sect.~\ref{sec:gyre}) for the $16.9\,\msun$ 3D MHD merger product and the $17.4\,\msun$ single star at different points along their MS evolution. As expected from the asymptotic period spacing $\Pi_{\ell}$, the mean values of the PSPs for the two models are relatively similar. At $X_{\mathrm{c,\,single}} = 0.61$, we see a quasi-periodic departure of the PSP from its asymptotic behaviour ($\Delta P_{n} = $ constant). This quasi-periodic variation is caused by the steep chemical gradient left behind by the receding convective core. This chemical gradient is deduced from the H profiles shown in Fig.~\ref{fig:mhd_props_and_comps_evolution} and causes a sharp variation in the BV frequency $\Tilde{N}$. We label this sharp variation in $\Tilde{N}$ as `peak 1' in the upper-left panel of Fig.~\ref{fig:mhd_props_and_comps_evolution}. The width of this peak increases with decreasing $X_{\mathrm{c}}$ because of the receding convective core and the subsequent extension of the near-core region with a strong chemical composition gradient. The abrupt change in $\Tilde{N}$, as seen from the peak, can trap g modes in the peak region, and it is this mode trapping that is responsible for the quasi-periodic deviations from the asymptotic PSP behaviour as observed in MS g-mode pulsators \citep[\eg][]{Michielsen2021}. Generally, the more the convective core recedes, the higher and broader the $\Tilde{N}$ profile (see \citealt{Aerts2021}) and the higher the probability of mode trapping. This is apparent from the PSPs for our models shown in Fig.~\ref{fig:psp_evolution}. From both theory \citep{Miglio2008, Cunha2015, Cunha2019, Cunha2024} and observations \citep{Moravveji2015, Moravveji_etal2016, Michielsen2021,Michielsen2023}, we know that the occurrence rate of dips in the PSP is related to the location and width of the sharp variation in $\Tilde{N}$ within the g-mode cavity. Because of the similar location of the sharp $\Tilde{N}$-variations (peak 1) with respect to the g-mode cavity in the merger product and the genuine single star, the periodicity in the PSP variations is relatively similar. However, there appears to be an additional component to this variation, most clearly seen at $X_{\mathrm{c,\,merger}} = 0.60$, in the PSPs of the merger product. We attribute this to the presence of a second sharp variation in the $\Tilde{N}$ profile of the merger product, labelled as `peak 2' in the upper-left panel of Fig.~\ref{fig:mhd_props_and_comps_evolution}. This peak is a remnant of the transient convective core described in Sect.~\ref{sec:methods:mhd}, which has left behind a strong chemical gradient at the location of its largest extent. Narrow peaks such as this can perturb and even trap g modes as long as their radial extent is smaller than or comparable to the local wavelength of the modes in question \citep{Cunha2020}. In these cases, the modes experience an abrupt change in $\Tilde{N}$ (as opposed to a smoothly-varying $\Tilde{N}$), which causes them to be perturbed. We can verify whether this is the case in our model: for modes with a period around $10$ days (these are the shortest-wavelength modes considered here), the local wavelengths are $\lambda_{\mathrm{local}} \approx 0.03\,\rsun$ and $\lambda_{\mathrm{local}} \approx 0.15\,\rsun$ at $X_{\mathrm{c,\,merger}} = 0.60$ and $X_{\mathrm{c,\,merger}} = 0.01$, respectively. The full-width-half-maxima (FWHM), which we use as a proxy for the radial extent of the Gaussian-like shape of the second peak, are $\mathrm{FWHM}_{\mathrm{peak\,2}} \approx 0.02\,\rsun$ and $\mathrm{FWHM}_{\mathrm{peak\,2}} \approx 0.08\,\rsun$ for $X_{\mathrm{c,\,merger}} = 0.60$ and $X_{\mathrm{c,\,merger}} = 0.01$, respectively. This shows that we expect this extra peak in the BV frequency profiles of the merger product to affect the g modes since $\lambda_{\mathrm{local}} \gtrsim \mathrm{FWHM}_{\mathrm{peak\,2}}$.

The amplitude of the PSP variation, that is, the departure of $\Delta P_{n}$ from the asymptotic value $\Pi_{\ell}$, depends on the sharpness, height, and width of the variation in $\Tilde{N}$ \citep{Miglio2008, Cunha2015,Cunha2019,Cunha2024}. The $\Tilde{N}$ profiles of 26 Slowly Pulsating B-type (SPB) pulsators observed by \textit{Kepler} and modelled by \citet{Pedersen2021} show a large variety in their sharpness, height, and width. Here, we see that the PSP variations' amplitudes are similar for the two models, albeit slightly higher for the merger product. The near-core peaks (peak 1) in the BV frequency profiles of the genuine single star are sharper than those for the merger product, while the extra peak in the BV frequency profile of the merger product introduces additional quasi-periodic variability with a different occurrence rate. Despite the similar magnitude of the amplitudes, their differences are typically larger than the observational uncertainties on $\Delta P_{n}$ of the B-type stars with the best asteroseismic measurements described by the uncertainty $\sigma_{\Delta P}$.\\

We now repeat the comparison above in the presence of rotation. As argued in Sects.~\ref{sec:mesa} and \ref{sec:gyre}, we include the effects of rotation (more specifically, the Coriolis force) at the level of the pulsation equations. Figure~\ref{fig:psp_rotation} shows the PSPs for prograde ($m > 0$) and retrograde ($m < 0$) sectoral ($\ell = |m|$) g modes predicted for the merger product and genuine single star at $X_{\mathrm{c,\,merger}} = 0.50$ and $X_{\mathrm{c,\,single}} = 0.51$, respectively. We consider rotation rates of $\Omega/\Omega_{\mathrm{c}} = 0.10\text{--}0.60$, with 
\begin{equation}
    \Omega_{\mathrm{c}} = \sqrt{GM_{\star}/R_{\mathrm{eq}}^{3}} \simeq \sqrt{8GM_{\star}/27R^3_{\star}}
\end{equation}
the Roche critical angular rotation frequency \citep{Maeder2009}, $G$ the gravitational constant, and $R_{\mathrm{eq}}$ the stellar radius at the equator of a rotationally deformed star. As expected from \citet{Bouabid2013}, the period spacings $\Delta P_{n}$ between the prograde modes become smaller with longer oscillation periods and with higher rotation rates, resulting in a negative slope of the PSP, while the retrograde PSPs have positive slopes, in agreement with observations of SPB pulsators \citep{Papics2014, Papics2015, Papics2017, Szewczuk2018, Szewczuk2021, Pedersen2021}. Overall, we see that the PSPs of the merger product and genuine single star have similar morphologies, that is, they follow the same trends. For prograde modes, the largest differences between the PSPs of the two models are found at shorter periods, but the difference remains clear at longer periods. The main differences in PSP variability are caused by the slightly different positions of the BV peaks. Whereas the average value $\Delta P_{n}$ for the merger product is lower than that for the genuine single star at $X_{\mathrm{c,\,merger}} = 0.50$, the opposite is true for prograde modes with longer periods when we take into account the effects of rotation. This is true for all rotation rates considered here. This can be seen in Fig.~\ref{fig:psp_rotation_diffs}, where we show the estimated differences the larger-period end\footnote{We only consider modes with periods $P_{n}$ satisfying $\left(P_{n} - P_{n}^{\mathrm{min}}\right)/\left(P_{n}^{\mathrm{max}} - P_{n}^{\mathrm{min}}\right) > 0.75$, with $P_{n}^{\mathrm{min}}$ and $P_{n}^{\mathrm{max}}$ the minimum and maximum period shown in Fig.~\ref{fig:psp_rotation}.} of the prograde-mode PSPs. We use a fit through the PSPs to make these estimations because it is not possible to do a one-to-one comparison between the modes of these models. Since we computed these PSPs under the TAR with \texttt{GYRE}, it would seem straightforward to fit them with the asymptotic period spacing relation under the TAR \citep[see, \eg Eq.~4 in][]{Bouabid2013}. However, such fits do not return satisfying fitting results when the PSPs deviate strongly from the otherwise smooth asymptotic behaviour under the TAR \citep{VanReeth2016}, as is the case in our PSPs. Hence, we fitted the PSPs in Fig.~\ref{fig:psp_rotation} with quadratic functions instead. Comparing the estimated differences with $\sigma_{\Delta P}$, we see that the differences $\Delta P_{n}$ could technically be observed for most prograde dipole ($\ell = 1$) modes and some prograde quadrupole ($\ell = 2$) modes in stars with $\Omega/\Omega_{\mathrm{c}} \leq 0.20$.

The PSPs for retrograde modes become `stretched' towards longer periods, accentuating the differences in PSP variability between the merger product and genuine single star even more. Lastly, we note the presence of relatively deep dips in the period spacing patterns of both the merger product and genuine single star for different mode morphologies and rotation rates. We do not find such deep dips in the non-rotating case for $\ell=1$ modes in the period range shown in Fig.~\ref{fig:psp_evolution}, but they are present at longer periods (higher radial order $n_{\mathrm{g}}$) and in the $\ell = 2$ modes (see Fig.~\ref{fig:psp_evolution_l2}). Closer inspection shows that these deep dips are caused by the coupling between g modes in the main inner g-mode cavity and those in the subsurface g-mode cavity. This is reminiscent of the g-g-mode coupling described in \citet{Unno1989book} and \citet{Henneco2024b}.

\begin{figure}[h!]
    \centering
    \resizebox{0.9\hsize}{!}{\includegraphics{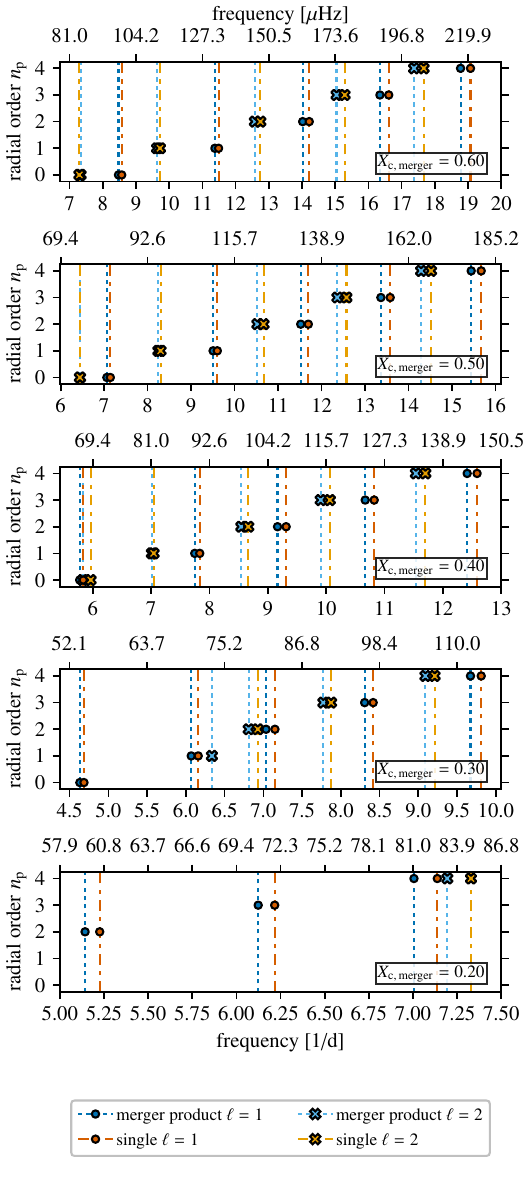}}
    \caption{Frequencies and radial orders $n_{\mathrm{p}}$ of $\ell = 1$ and $\ell = 2$ p modes with $n_{\mathrm{p}} \leq 4$ for the $16.9\,\msun$ 3D MHD merger product (blue and light-blue markers) and the $17.4\,\msun$ genuine single star (red and orange markers) at different evolutionary stages, in the absence of rotation. The models at $X_{\mathrm{c}} = 0.01$ and $X_{\mathrm{c}} = 0.10$ are absent because of the lack of pure p modes at these evolutionary stages. The dot and cross markers show the value of the radial order $n_{\mathrm{p}}$ for the $\ell = 1$ and $\ell = 2$ modes, respectively. The dotted and dash-dotted lines are drawn to accentuate the frequency differences and improve the legibility of the frequency values.}
    \label{fig:pmodes_nonrot}
\end{figure}

\begin{figure}
    \centering
    \resizebox{0.9\hsize}{!}{\includegraphics{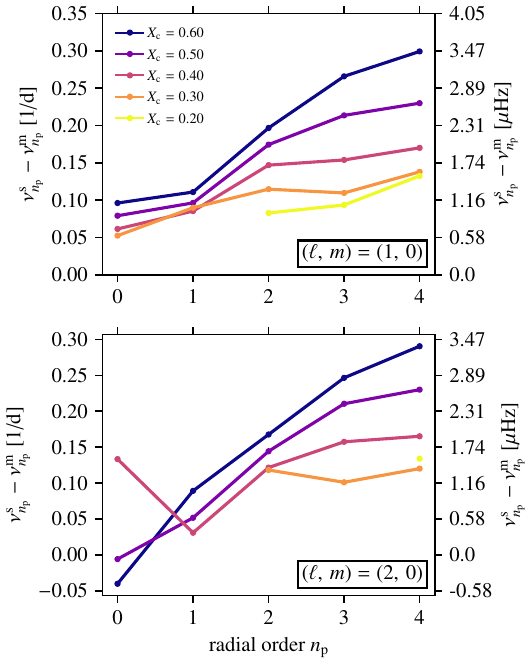}}
    \caption{Absolute differences between the $17.4\,\msun$ genuine single star's p-mode frequencies $\nu_{n_\mathrm{p}}^{\mathrm{s}}$ and the $16.9\,\msun$ 3D MHD merger product's p-mode frequencies $\nu_{n_\mathrm{p}}^{\mathrm{m}}$ per radial order $n_\mathrm{p}$, in the absence of rotation. The $(\ell,\,m) = (1,\,0)$ and  $(\ell,\,m) = (2,\,0)$ modes are shown in the upper and lower panel, respectively. The lines are colour-coded according to the merger product's central hydrogen fraction $X_{\mathrm{c}}$.}
    \label{fig:pmodes_nonrot_differences}
\end{figure}

\subsubsection{Low-order pressure modes}\label{sec:paper3_pmodes}

Figure \ref{fig:pmodes_nonrot} shows the frequencies of the low-radial order $\ell = 1$ and $\ell = 2$ p modes with radial orders $n_{\mathrm{p}} \leq 4$ and without rotation. Pressure modes with higher radial orders are not observed in $\beta$~Cep stars \citep{Fritzewski2024b}. We compare the predicted p modes for the $16.9\,\msun$ 3D MHD merger product and the $17.4\,\msun$ genuine single star at different stages during their MS evolution. We find no pure p modes, that is, modes with $n_{\mathrm{g}} = 0$ for the models at $X_{\mathrm{c,\,merger}} = 0.01$ and $X_{\mathrm{c,\,merger}} = 0.10$ because the structures of the g- and p-mode cavities start to overlap more in frequency, leading to mode mixing \citep{Unno1989book}, hence, we leave these evolutionary stages out. Because of their similar p-mode cavities, the p modes in the merger product and genuine single star span a similar frequency range. It is also clear from Fig.~\ref{fig:pmodes_nonrot} and Fig.~\ref{fig:pmodes_nonrot_differences} that the p-mode frequencies of the genuine single star are higher by at most 0.3 cycles per day ($3.5\,\mu$Hz). Furthermore, this frequency difference increases with radial order $n_{\mathrm{p}}$ and decreases with MS age (Fig.~\ref{fig:pmodes_nonrot_differences}). Even the smallest frequency difference, which we predict for the fundamental ($n_{\mathrm{p}}=0$) $\ell = 1$ mode at $X_{\mathrm{c}}=0.30$, has a relative value of ${\sim}10\%$. This is 1000 times larger than the observed relative p-mode frequency uncertainty of ${\sim}0.01\%$ reported in \citet{Aerts2019} and the absolute p-mode frequency uncertainty of $\sigma_{\nu}^{\mathrm{p}} \simeq 0.01\mu$Hz found for the prototypical $\beta$~Cep star HD~129929 by \citet{Aerts2003,Aerts2004}. Overall, the merger product's p modes are more closely spaced than the genuine single star's. 

The difference in p-mode frequencies can be explained by the merger product's lower mean density in its envelope, shown in Fig.~\ref{fig:envelope_rho_cs}. This relates back to the generic principle that since the merger product has the same luminosity, effective temperature, and radius as a more massive genuine single star, its mean envelope molecular weight should be higher and its mean density should be lower, as is indeed the case. The decrease of the difference in p-mode frequencies between the two models with MS age can be attributed to the fact that the merger product's and genuine single star's mean envelope density become similar. Since p modes with higher radial order have more nodes in the region of the star when the merger product and genuine single star differ significantly, they are more sensitive to these differences than those with fewer radial nodes. Just as for g modes, we see that the differences between the merger product's and genuine single star's predicted asteroseismic characteristics are largest when the stars are younger.\\

Analogue to the g modes, we consider the effect of rotation, more specifically the Coriolis force. We treat it perturbatively up to first order in the rotation frequency, which is a good approximation for the p modes for slow and modest rotators. Its influence on the p modes and on the frequency difference found in the absence of rotation can be found in Appendix~\ref{app:rotating_pmodes}. We find that, at least with perturbative implementation of the effect of the Coriolis force, the differences in p-mode frequencies between the merger product and genuine single star are similar to those in the absence of rotation.

\subsection{Comparison with and between 1D merger methods}

\begin{figure}
    \centering
    \resizebox{0.9\hsize}{!}{\includegraphics{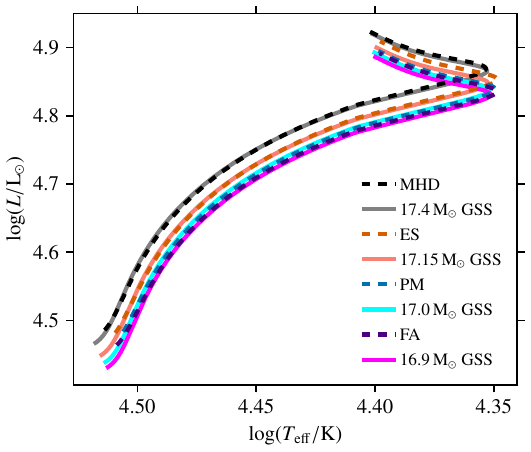}}
    \caption{HRD with the evolutionary tracks of the $16.9\,\msun$ merger product computed with the 3D MHD simulation (MHD, black dashed line), entropy sorting (ES, red dashed line), \texttt{PyMMAMS} (PM, blue dashed line), and fast accretion (FA, indigo dashed line). The corresponding $17.4\,\msun$, $17.15\,\msun$, $17.0\,\msun$, and $16.9\,\msun$ genuine single stars evolutionary tracks are drawn with grey, orange, cyan, and magenta solid lines, respectively.}
    \label{fig:hrd_merger_methods}
\end{figure}

\begin{figure}
    \centering
    \resizebox{0.9\hsize}{!}{\includegraphics{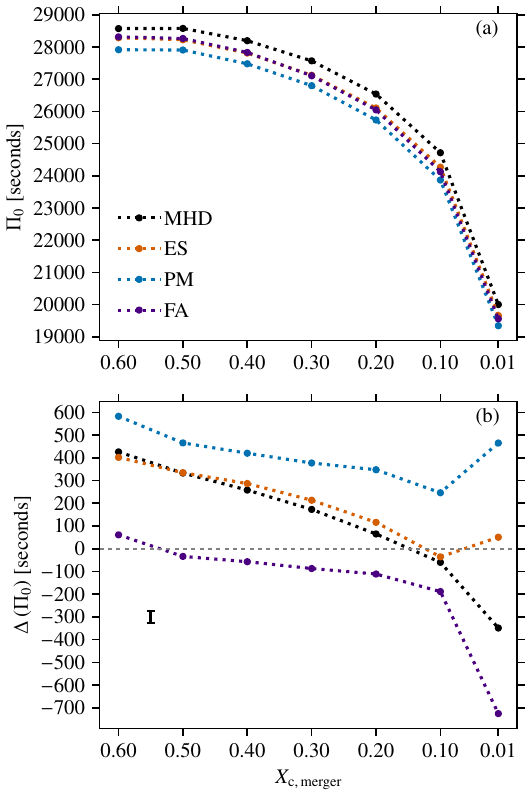}}
    \caption{Comparison between the buoyancy travel time $\Pi_{0}$ for the $16.9\,\msun$ 3D MHD (MHD, black line), entropy-sorted (ES, red line), \texttt{PyMMAMS} (PM, blue line), and fast accretion (FA, indigo line) merger products (Panel a), and the absolute differences $\Delta (\Pi_{0}) = \Pi_{0,\,\mathrm{single}} - \Pi_{0,\,\mathrm{merger}}$ with their respective genuine single star models (Panel b). The error bar in Panel (b) shows the $\sigma_{\Delta P}$.}
    \label{fig:Pi_vs_Xc_method_compare}
\end{figure}

We now investigate if and how the predictions made in Sect.~\ref{sec:results_astero_compare} differ when we use 1D merger prescriptions. As mentioned in Sect.~\ref{sec:paper3_introduction}, we consider three commonly used 1D merger prescriptions: entropy sorting, entropic variable sorting with shock heating (\texttt{PyMMAMS}), and fast accretion (see Sect.~\ref{sec:mergermethods}). The HRD in Fig.~\ref{fig:hrd_merger_methods} shows evolutionary tracks of the merger products acquired with all four methods (the 3D MHD simulation and the three 1D methods). We see that the tracks do not coincide in the HRD, which results in each merger product approximately overlapping with a genuine single star of a different mass. As detailed at the beginning of Sect.~\ref{sec:paper3_results}, the track for the entropy-sorted model comes closest to that of the 3D MHD model; its corresponding genuine single star has a mass of $17.15\,\msun$. Using the entropy sorting method, we under-predict $\Pi_{0}$ by at most $450\,$s in comparison to the 3D MHD model. The error one would make by using entropy sorting instead of the 3D MHD model would thus be around $9\times\sigma_{\Delta P}$. In other words, the error we make by using entropy sorting is of the same order as the difference in $\Pi_{0}$ between the 3D MHD merger product model and its corresponding genuine single star (see Sect.~\ref{sec:paper3_gmodes}). With entropy sorting, we under-predict this $\Pi_{0}$ difference between the merger product and its corresponding genuine single star compared to the 3D MHD merger product (Fig.~\ref{fig:Pi_vs_Xc_method_compare}b). This difference has a median value of $213\,$s and lies in the interval $[-36;\,402]\,$s. 

The HRD tracks of the \texttt{PyMMAMS} and fast accretion merger products also lie below that of the 3D merger product, corresponding to genuine single-star models with masses of $17.0\,\msun$ and $16.9\,\msun$, respectively (Fig.~\ref{fig:hrd_merger_methods}). With both methods, we under-predict $\Pi_{0}$ compared to the 3D MHD merger product by at least $657\,$s and $257\,$s, and at most $850\,$s and $590\,$s for the \texttt{PyMMAMS} and fast-accretion merger product, respectively. The resulting differences in $\Pi_{\ell}$ are significantly larger than $\sigma_{\Delta P}$. With the \texttt{PyMMAMS} method, we over-predict the difference in $\Pi_{0}$ compared to its corresponding genuine single star. The median value of this difference is $420\,$s and it lies in the interval $[246;\,583]\,$s. With the fast accretion method, we under-predict the difference in $\Pi_{0}$ between the merger product and the genuine single star. This absolute difference has a median value of $-87\,$s and lies in the interval $[-726;\,61]\,$s. The $\Pi_{0}$ difference between all merger products and their genuine single-star counterparts increases abruptly in absolute value at $X_{\mathrm{c,\,merger}} = 0.01$. We attribute this to the fact that there is a noticeable difference between the HRD tracks around the Henyey hook (the point where $\log T_{\mathrm{eff}}$ starts increasing; see Fig.~\ref{fig:hrd_merger_methods}). The overall differences in the $\Pi_{0}$ values and differences in $\Pi_{0}$ between the merger products obtained through 1D methods and their corresponding genuine single stars can be explained by the differences in the merger products' BV profiles compared to the 3D MHD merger product. We highlight these differences in the following sections.

\subsubsection{Entropy-sorted merger product}\label{sec:ES_results}

\begin{figure*}
\centering
  \includegraphics[width=16cm]{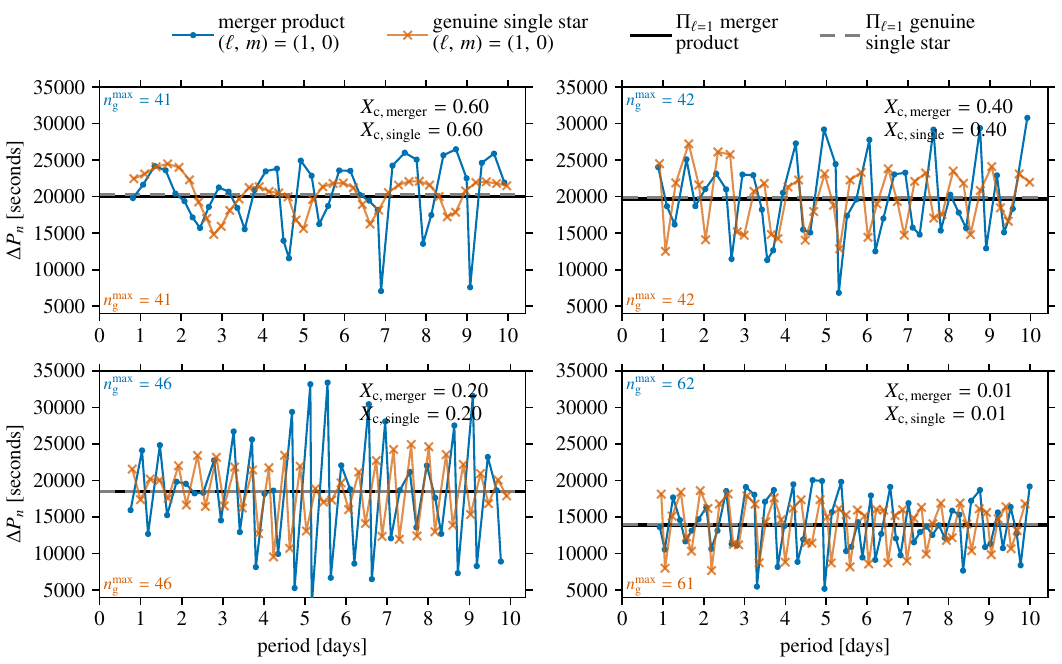}
     \caption{Same as Fig.~\ref{fig:psp_evolution}, now for the $16.9\,\msun$ entropy sorted merger product (blue solid lines, dot markers) and the $17.15\,\msun$ genuine single star (orange solid lines, cross markers).}
     \label{fig:ES_psp_evolution}
\end{figure*}

\begin{figure}
    \centering
    \resizebox{0.90\hsize}{!}{\includegraphics{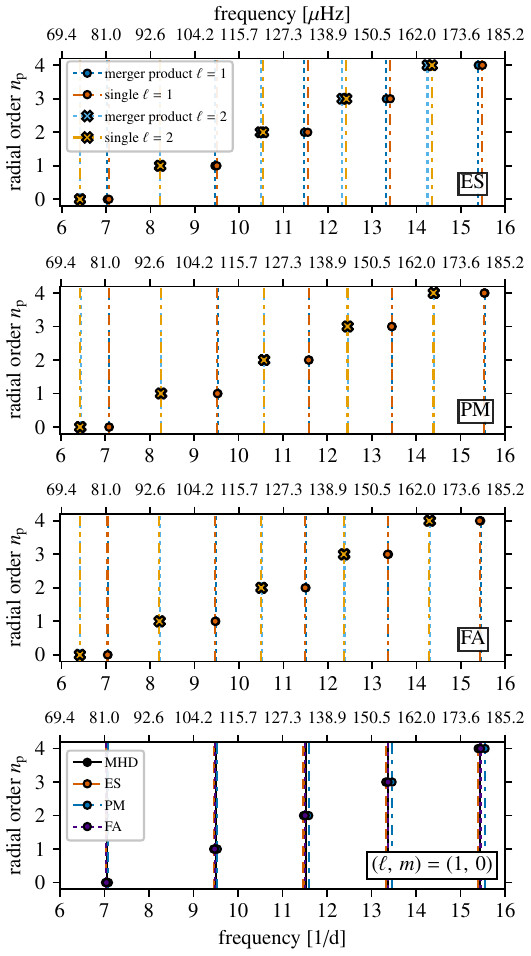}}
    \caption{Mode frequencies and radial orders $n_{\mathrm{p}}$ for $(\ell,\,m) = (1,\,0)$ and $(\ell,\,m) = (2,\,0)$ p modes in the absence of rotation at $X_{\mathrm{c,\,merger}} = 0.50$ for the entropy-sorted (ES), \texttt{PyMMAMS} (PM), and fast accretion (FA) merger product models and their corresponding genuine single-star models. In the top three panels, the colour, line, and marker conventions are the same as in Fig.~\ref{fig:pmodes_nonrot}. The bottom panel shows the $(\ell,\,m) = (1,\,0)$ p mode frequencies for the 3D MHD, entropy-sorted, \texttt{PyMMAMS}, and fast accretion models together.}
    \label{fig:pmodes_nonrot_method_compare}
\end{figure}

\begin{figure}
    \centering
    \resizebox{0.90\hsize}{!}{\includegraphics{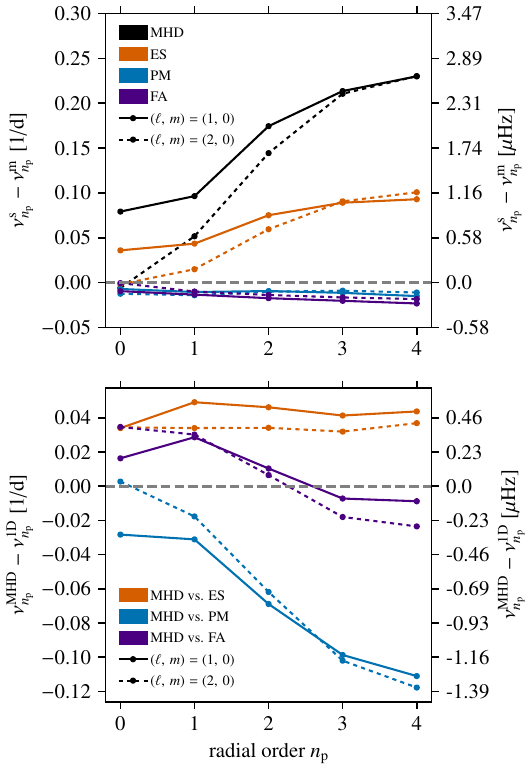}}
    \caption{Absolute differences between the $17.4\,\msun$, $17.15\,\msun$, $17.0\,\msun$, and $16.9\,\msun$ genuine single star's p mode frequency $\nu_{n_\mathrm{p}}^{\mathrm{s}}$ and the $16.9\,\msun$ 3D MHD (MHD), entropy-sorted (ES), \texttt{PyMMAMS} (PM), and fast-accretion (FA) merger product's frequency $\nu_{n_\mathrm{p}}^{\mathrm{m}}$ per radial order $n_\mathrm{p}$, in the absence of rotation (top panel). The bottom panel shows the difference in p-mode frequencies per radial order for p modes computed with the 3D MHD merger model as input in \texttt{GYRE} and those computed with the respective 1D models used as input.}
    \label{fig:pmodes_nonrot_differences_method_compare}
\end{figure}

From Fig.~\ref{fig:es_props_and_comps_evolution}, we see that the near-core region of the entropy-sorted merger product model is enriched in He, but less strongly and over a smaller radial extent than what we found in the model for the 3D MHD merger product. Contrary to the 3D MHD merger product, the entropy-sorting method does not lead to the He-rich core of the primary star ending up in a shell around the secondary's core. The He enrichment originates from the secondary’s core -- also enriched in He, though less strongly than the primary’s -- which forms a layer around the merger product’s core. Additionally, transient convective zones emerge around the core-envelope boundary during the thermal relaxation phase, mixing He-rich core material into the near-core region. Contrary to the 3D MHD model, the entropy sorted model does not have a transient convective core before starting its MS evolution. However, its BV frequency profile contains multiple Gaussian-like peaks (Fig.~\ref{fig:es_props_and_comps_evolution}) due to the staircase-like structure of the chemical composition profile, which has two origins. The first origin is the entropy-sorting method itself, which can result in large jumps in the chemical composition between neighbouring layers of the star. The second origin is the appearance of short-lived, numerically unstable convection zones before the merger product settles on the MS.

Looking at the predicted PSPs for the entropy-sorted merger model and its $17.15\,\msun$ genuine single-star counterpart in Fig.~\ref{fig:ES_psp_evolution}, we see a similar trend in the asymptotic period spacings $\Pi_{\ell}$ as in the case of the 3D MHD merger product. The values of $\Pi_{\ell}$ for the merger product and genuine single star differ by more than $\sigma_{\Delta P}$, yet these differences are, in general, somewhat higher (except at $X_{\mathrm{c,\,merger}}=0.60$) than what we predict for the 3D MHD merger product. The variability in the PSPs of the entropy-sorted model is seemingly more chaotic than that for the 3D MHD merger product because the modes are affected by multiple peaks in the $\Tilde{N}$-profile. The amplitude of the variability is also higher for the entropy-sorted model.

From Figs.~\ref{fig:pmodes_nonrot_method_compare} and \ref{fig:pmodes_nonrot_differences_method_compare}, we see that the predicted p modes for the entropy-sorted merger product model behave qualitatively similarly to those predicted for the 3D MHD model. However, we under-predict the differences in $\ell=1$ and $\ell=2$ p-mode frequencies between the merger product and genuine single star by up to 0.14\,cycles/day (1.62\,$\mu$Hz) compared to the 3D MHD model, which is more than two orders of magnitude larger than $\sigma_{\nu}^{\mathrm{p}} \simeq 0.01\,\mu$Hz \citep{Aerts2003,Aerts2004}. Furthermore, the frequency error we make by using entropy sorting instead of the 3D MHD model, shown in the bottom panel of Fig.~\ref{fig:pmodes_nonrot_differences}, is on the order of 0.03--0.05\,cycles/day (0.35--0.58\,$\mu$Hz) and is also significantly larger than  $\sigma_{\nu}^{\mathrm{p}}$. We attribute these frequency shifts to the different chemical structures (Fig.~\ref{fig:es_props_and_comps_evolution}) the two merger product models have in their p-mode cavities (see Sect.~\ref{sec:paper3_pmodes}). 

\subsubsection{\texttt{PyMMAMS} merger product}
\begin{figure*}
\centering
  \includegraphics[width=16cm]{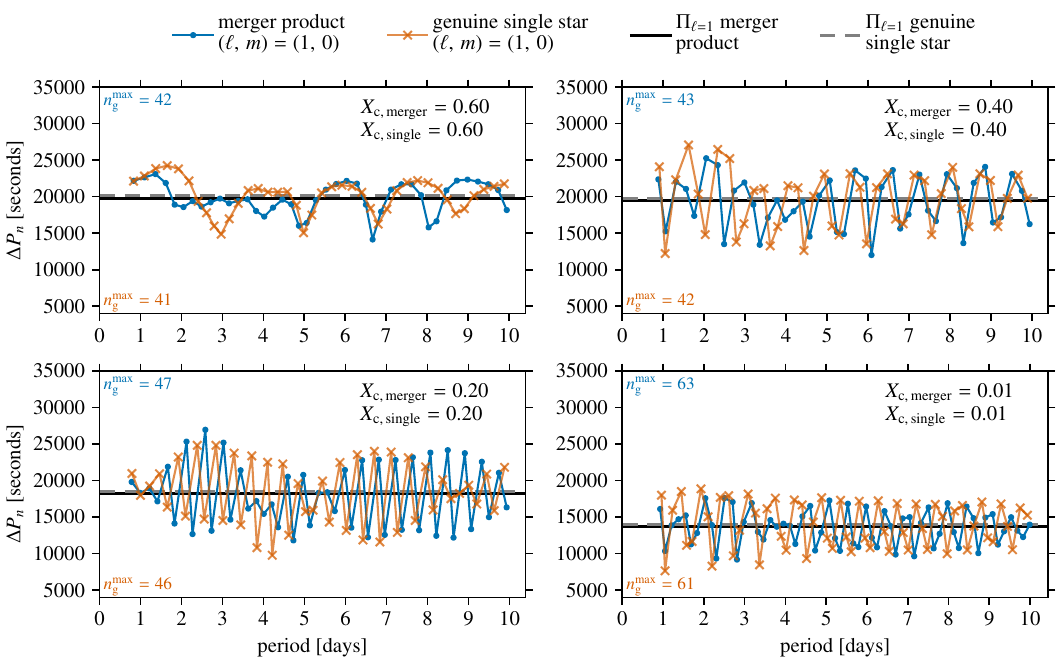}
     \caption{Same as Fig.~\ref{fig:psp_evolution}, now for the $16.9\,\msun$ \texttt{PyMMAMS} merger product (blue solid lines, dot markers) and the $17.0\,\msun$ genuine single star (orange solid lines, cross markers).}
     \label{fig:PM_psp_evolution}
\end{figure*}

Thanks to the addition of shock heating in the \texttt{PyMMAMS} prescription, it performs better in reproducing the overall merger product structure expected from the 3D MHD simulation. Contrary to the entropy sorted model, the \texttt{PyMMAMS} prescription results in the secondary's core sinking to the centre of the merger product and the primary's core forming a shell around it (see Fig.~\ref{fig:pm_props_and_comps_evolution}). However, we find that by comparing the chemical composition profiles for the 3D MHD and\texttt{PyMMAMS} merger products (Figs.~\ref{fig:mhd_props_and_comps_evolution} and \ref{fig:pm_props_and_comps_evolution}), the He-enrichment of the merger product's envelope is limited to the near-core region (up to $r \approx 1.8\,\rsun$ at $X_{\mathrm{c}}=0.60$) in the \texttt{PyMMAMS} model, whereas the enrichment extends further out to $r\approx3.0\,\rsun$ in the 3D MHD model. A second difference between the 3D MHD, entropy-sorted, and \texttt{PyMMAMS} models is the lack of an extended transient convective core during the merger product's thermal relaxation phase before settling back on the MS in the latter model. The extra peak in the $\Tilde{N}$ profile (peak 2), present in the 3D MHD, is missing in the \texttt{PyMMAMS} model. As a result, the BV frequency profiles for this merger product and its corresponding $17.0\,\msun$ genuine single star model look almost identical (Fig.~\ref{fig:pm_props_and_comps_evolution}). Only the radial extent of the BV frequency peak of the merger product is larger than in the genuine single-star model because of the He-enrichment in the near-core region. The lack of an additional peak in the BV frequency profile of the \texttt{PyMMAMS} model causes there to be no additional variability in the merger model's PSPs (Fig.~\ref{fig:PM_psp_evolution}). The two PSPs have roughly the same quasi-periodic behaviour, albeit with a relatively small phase shift likely caused by the more extended BV frequency peak of the \texttt{PyMMAMS} merger product. However, the phase shift and difference in $\Delta P_{n}$ are still larger than the currently best uncertainty of g-mode periods $\sigma_{P}^{\mathrm{g}} \simeq 2.56\times 10^{-4}\,$days \citep{Moravveji2015} and $\sigma_{\Delta P}$, respectively.

Because of the virtually identical chemical composition profiles and p-mode cavities (Fig.~\ref{fig:pm_props_and_comps_evolution}), we find that the p-mode frequency absolute differences between the \texttt{PyMMAMS} merger model and its corresponding genuine single star are negative and\linebreak ${\leq}0.02\,$cycles/day (${\leq}0.23\,\mu$Hz), shown in the top panel of Fig.~\ref{fig:pmodes_nonrot_differences}. This is an order of magnitude larger than $\sigma_{\nu}^{\mathrm{p}}$. The p-mode frequency error compared to the 3D MHD model (bottom panel of Fig.~\ref{fig:pmodes_nonrot_differences}) is on the order of $0.1\,$cycles/day (1.16$\,\mu$Hz), several orders of magnitude larger than $\sigma_{\nu}^{\mathrm{p}}$.

\subsubsection{Fast accretion merger product}\label{sec:FA_results}
\begin{figure*}
\centering
  \includegraphics[width=16cm]{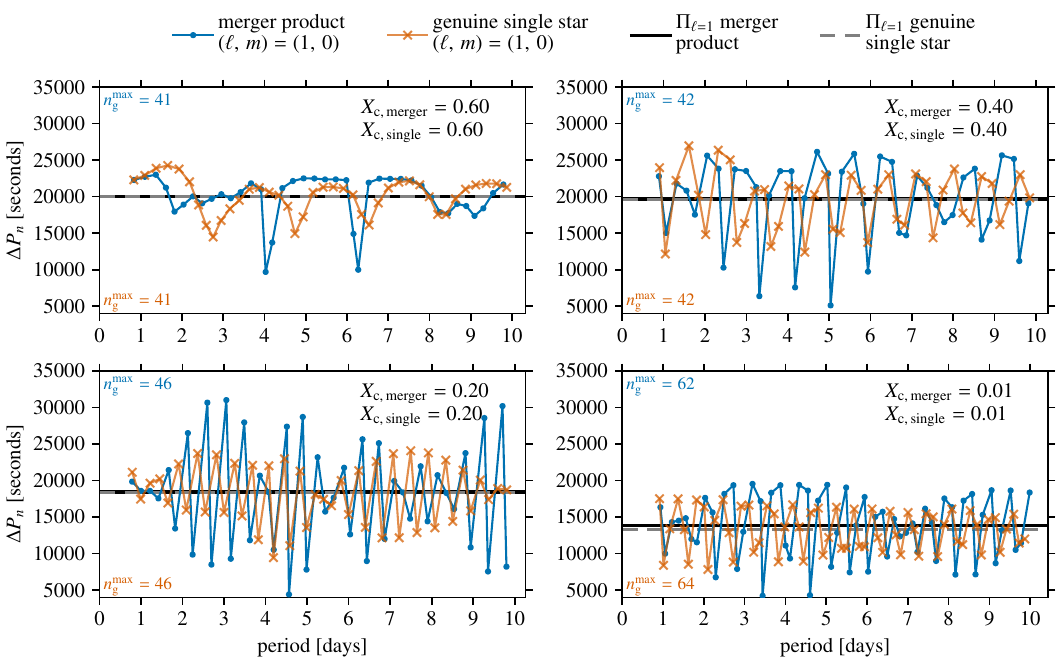}
     \caption{Same as Fig.~\ref{fig:psp_evolution}, now for the $16.9\,\msun$ fast accretion merger product (blue solid lines, dot markers) and the $16.9\,\msun$ genuine single star (orange solid lines, cross markers).}
     \label{fig:FA_psp_evolution}
\end{figure*}

Although the fast accretion method has proven to be sufficient in reproducing low-mass \citep{Rui2021} and massive \citep{Henneco2024b} post-MS merger products, it does not perform well for MS merger products. For massive post-MS merger products, the sub-thermal-timescale accretion onto a blue Hertzsprung-gap star emulating the merger leads to the formation of long-lived blue supergiant stars with distinct structures. In the MS case, we see that the fast accretion phase leads to a transient convective core, yet this core has a smaller extent than what we found in the 3D MHD and entropy-sorted model. This transient convective core and the subsequent receding rejuvenated convective core leave an imprint in the chemical composition profile, which results in double-peaked BV frequency profile in the near-core region (Fig.~\ref{fig:fa_props_and_comps_evolution}), similar to the one found in \citet{Wagg2024}. From \citet{Henneco2024b}, we expect that the extent of this transient convective core depends on the mass added to the primary star, the accretion timescale, and the semi-convective efficiency \citep{Braun1995}. We see from Fig.~\ref{fig:FA_psp_evolution} that the second BV frequency peak influences the PSP variability mostly in the number of modes found in the PSP dips and the amplitude, which are both higher for the merger product.

On the level of the p modes, we see similar behaviour as for the \texttt{PyMMAMS} model. Since we assumed that the composition of the accreted material is that of the surface of the accreting star, the envelope of the merger product is not enriched in helium. Because of the small radial extent of the transient convective core during the merger procedure, the p-mode cavity is virtually identical to the $16.9\,\msun$ genuine single star's p-mode cavity. The frequency differences between the merger product and genuine single star are at most $0.025\,$cycles/day (0.289$\,\mu$Hz). The frequency error compared to the 3D MHD merger product's p modes is in the interval $[-0.025;\,0.036]\,$cycles/day ($[-0.289;\,0.417]\,\mu$Hz), but is larger in absolute value than $\sigma_{\nu}^{\mathrm{p}}$ for all radial orders (bottom panel of Fig.~\ref{fig:pmodes_nonrot_differences}).

\subsubsection{Potential improvements to 1D merger prescriptions}\label{paper3:improvements}
Since none of the 1D merger prescription models is able to reproduce the interior structure and asteroseismic predictions of the 3D MHD model, we briefly discuss some ways in which these methods could potentially be improved. Both the mean asymptotic period spacing values and PSP variability morphology of the 3D MHD merger product model are not reproduced well with the fast accretion method. A potential improvement to this method would be to abandon the assumption that the chemical composition of the accreted material is the same as that of the accretor's surface and instead accrete the full chemical composition of the secondary star. However, this has the drawback that one needs to make assumptions about the mixing of this material in the accretor's envelope. 

The \texttt{PyMMAMS} model performed worse than the entropy-sorted model, but has the highest potential for improvement, especially since it predicts the correct overall chemical structure of the merger product (secondary core in the centre, primary core around it). As mentioned in Sect.~\ref{sec:methods_pymmams}, the main difference between entropy-sorting and \texttt{PyMMAMS} is the inclusion of shock heating in the latter. Evidently, the shock heating, calibrated on more energetic head-on collisions, does not lead to a satisfactory reproduction of the 3D MHD model. However, provided that more 3D binary merger simulations become available, they could be used to calibrate the shock heating prescription in \texttt{PyMMAMS} to better reproduce the chemical composition profiles resulting from these slower, less energetic binary inspiral mergers (Heller et al., in prep.). In the limit where a calibrated \texttt{PyMMAMS} version reproduces a model with the exact same chemical composition profiles as the 3D MHD model, we expect the subsequent evolution and asteroseismic signals to be the same.

\subsection{Discussion}
\begin{figure}[h!]
    \centering
    \resizebox{0.9\hsize}{!}{\includegraphics{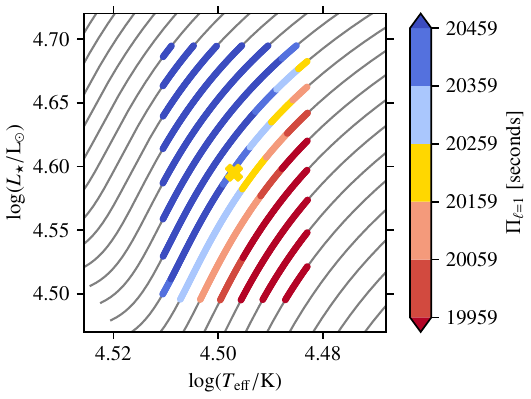}}
    \caption{Values of $\Pi_{\ell=1}=\Pi_{1}$ for the 3D MHD merger product at $X_{\mathrm{c}} = 0.50$ (cross marker, $\Pi_{1}^{\mathrm{merger}} = 20209\,$s) and genuine single stars (dot markers) in a typical observational uncertainty box of $\log\left(L_{\star}/\lsun\right) \pm 0.1\,$dex and $T_{\mathrm{eff}} \pm 1000\,$K. The grey lines are MS HRD tracks for genuine single stars with masses between $15.4\,\msun$ (bottom right) and $20.0\,\msun$ (top left). The boundaries of the colourbar are set by $\Pi_{1}^{\mathrm{merger}} \pm K\sigma_{\Delta P}$, with $K = 1,\,2,\,3$.}
    \label{fig:hrd_Pi1_shifts}
\end{figure}

In summary, the structural and chemical anomalies resulting from the merger lead, as shown in Sect.~\ref{sec:paper3_gmodes}, to different asymptotic period spacings $\Pi_{\ell}$ and hence mean PSP values for the merger product and genuine single star. The differences in the asymptotic period spacing are on the order of several $10\,$s to $100\,$s, which is larger than the current best uncertainties on g-mode period spacing patterns of $\sigma_{\Delta P} \simeq 50\,$s. However, as we have shown, this difference in the asymptotic period spacing changes as the star evolves (Fig.~\ref{fig:Pi_vs_Xc_matched}). Also, for prograde sectoral modes computed under the TAR, the differences in PSP values are smaller than without rotation, and the PSP values are, on average, higher for the merger product than for the genuine single star, whereas the opposite is true in the non-rotating case (at $X_{\mathrm{c}} = 0.50$). In other words, if observed period spacing patterns are available, their mean value likely does not unambiguously allow us to distinguish this type of merger product from genuine single stars.\\
We can look at this from another angle. In Fig.~\ref{fig:hrd_Pi1_shifts}, we show the asymptotic period spacing for $\ell=1$ modes $\Pi_{\ell=1} = \Pi_{1}$ for the 3D MHD merger product at $X_{\mathrm{c}} = 0.50$ and genuine single stars in a typical observational box centred around the HRD location of the merger product with a size of $\log\left(L_{\star}/\lsun\right) \pm 0.1\,$dex and $T_{\mathrm{eff}} \pm 1000\,$K. As readily seen in Sect.~\ref{sec:paper3_gmodes}, the difference in $\Pi_{1}$ is several times higher than $\sigma_{\Delta P} = 50\,$s for the merger product and its closest genuine single star model. Going to higher and lower masses, the absolute value of the difference in $\Pi_{1}$ quickly grows. In the limit of infinitely accurate measurements of $L_{\star}$ and $T_{\mathrm{eff}}$, we do expect the merger product to be an outlier in terms of $\Pi_{1}$. However, at slightly lower $\log T_{\mathrm{eff}}$ and similar $\log L_{\star}$ -- within the typical uncertainties on $L_{\star}$ and $T_{\mathrm{eff}}$ mentioned above -- we find genuine single-star models with $\Pi_{1}$ values comparable to that of the merger product. Thus, we would need highly accurate measurements of $L_{\star}$ and $T_{\mathrm{eff}}$ to distinguish this particular merger product; on the order of $0.001\,$dex in $\log T_{\mathrm{eff}}$. In other words, given today's typical uncertainties on $L_{\star}$ and $T_{\mathrm{eff}}$, we predict that the merger product is likely indistinguishable from genuine single stars based on its $\Pi_{1}$ value.

As described in Sect.~\ref{sec:paper3_pmodes}, frequencies of p modes are lower and more closely spaced in the merger product compared to the genuine single star, which we attribute to the merger product's lower mean density and sound speed in its p-mode cavity. These mean values converge with MS age, making the differences in p-mode frequencies smaller. Nevertheless, the differences are still larger than the current best relative uncertainties on observed p-mode frequencies of $0.01\%$ \citep[][Table\,1]{Aerts2019}. Additionally, the differences increase with increasing radial order because of the higher sensitivity of these modes to the chemical composition in the p-mode cavity. We also found that the frequency differences between the merger product and genuine single star are virtually insensitive to the rotation rate as a consequence of the first-order Ledoux perturbative approach used to include rotation on the level of the pulsation equations only. Especially at higher rotation rates, the effects of the centrifugal deformation of the star on p modes should not be ignored (see \citealt{AertsTkachenko2024} for an overview). Overall, our analysis shows that the differences in p-mode frequencies are on the order of $0.10\,$cycles/day ($1.16\,\mu$Hz) and depend strongly on the chemical composition of the p-mode cavity, which is the part of the star where the merger product deviates the most from its corresponding genuine single star.

We emphasise that a merger between two stars will leave a chemical signature in the merger product's envelope \citep{Glebbeek2013}. Exceptions are unevolved stars close to the zero-age main sequence (ZAMS), symmetrical binaries (\ie nearly equal mass stars that have evolved quasi-identical before the merger. This assumes that no significant mass transfer occurs during the contact phase of such systems), or mergers in which both progenitor cores end up in the product's core (with no chemical enrichment of the envelope of the result). This signature does depend on, for example, the mixing processes during and after the merger and will influence the asteroseismic signatures.

In our assessment of the performance of the three 1D merger methods, we find that none of the 1D methods can replicate the 3D MHD model structure. Entropy sorting performs best for this particular merger product even though this is likely coincidental. This method fails to reproduce the behaviour found in the 3D merger product model where the He-rich core of the primary star forms a layer around the secondary star's core, leading to strong and radially extended He-enrichment in the lower envelope. The \texttt{PyMMAMS} prescription does lead to an overall correct chemical structure for the merger product, yet, the radial extent of the He-enrichment of the envelope is less than in the 3D MHD model. By design, the fast accretion method is unable to reproduce the overall structure of the 3D MHD model, which is unlikely to improve by accreting material with a more realistic composition (see Sect.~\ref{paper3:improvements}). Because of the sensitivity of both g and p modes to the interior chemical structure, the asteroseismic predictions differ depending on which 1D merger prescription we used. More importantly, none of the asteroseismic predictions based on 1D merger prescription equilibrium models managed to reproduce those for the 3D MHD model within the observational uncertainties. In other words, all 1D methods introduce errors in the predicted $\Delta P_{n}$ and $\nu_{\mathrm{p}}$ that are larger than the current best observational uncertainties. Moreover, even with the best-performing method, entropy sorting, the error we make in $\Pi_{0}$ is of the same order as the predicted $\Pi_{0}$ differences between the 3D MHD merger model and its corresponding genuine single star. Therefore, it would be quite problematic if one were to use these 1D merger methods to create merger models for asteroseismic fitting and forward modelling purposes. Overall, as explained in Sect.~\ref{paper3:improvements}, the \texttt{PyMMAMS} model has the most potential to produce merger product models consistent with those obtained from 3D simulations on the condition that it is calibrated for binary mergers.

\section{Conclusions}\label{sec:paper3_dandc}

We find from the results presented in Sect.~\ref{sec:paper3_results} that even though the resulting product of a stellar merger between a $9\,\msun$ and a $8\,\msun$ MS star leads a seemingly `normal' MS life, its structure and composition are rather anomalous compared to genuine single stars. The $16.9\,\msun$ merger product obtained from the 3D MHD simulation of \citet{Schneider2019} has an abnormally high convective core radius and stellar radius for its respective mass. More specifically, throughout its MS evolution, the merger product's convective core and stellar radii are similar to those of a more massive $17.4\,\msun$ genuine single star with similar effective temperature and luminosity. Given that the merger product overlaps in the HRD with a more massive star, its $L_{\star}/M_{\star}$ is higher than that of genuine single stars. We have shown in Sect.~\ref{sec:paper3_results} that this can be explained by the He-enrichment of the envelope of the merger product and should be a generic feature in stars with chemically enriched envelopes.

The quasi-periodic variation found in the PSPs has the additional potential of singling out merger products from photometric light curves. The merger product's PSP variability contains approximately the same component as the genuine single star's, namely that caused by the strong BV frequency peak in their near-core regions. However, the extra peak in the merger product's BV frequency profile introduces a second component to this variability. Although it is harder to see at later MS stages, this second component in the merger product's variability distorts the somewhat regular behaviour of the variability that we would expect without the effect of the additional peak. This becomes clear when comparing the PSPs of the merger product and the genuine single star. A promising diagnostic for the presence of a merger product would be to look for such irregularities in observed PSPs. For that, mode identification (assigning $n$, $\ell$, and $m$ values to individual modes) is required, which currently lies within the realm from combined \textit{Gaia} and TESS space photometry \citep{Hey2024,Fritzewski2024b}. One could then look for and characterise components in the PSP variability, such as those predicted in this work. After this characterisation, tools such as those developed by \citet{Miglio2008} and \citet[][and references therein]{Cunha2024} could be used to link potential merger products' particular PSP variability to their internal structure. These tools are derived for non-rotating stars. So, before they can be used, the effect of the Coriolis force, which dominates the PSPs' deviation from their asymptotic consonant behaviour (\ie the slopes of the PSPs), needs to be modelled first using TAR-based PSP predictions. Various generalisations to the basic PSP predictions based on the TAR have been published, including the effects of differential rotation \citep{VanReeth2018}, small centrifugal deformation \citep{Henneco2021, Dhouib2021}, and an additional internal magnetic field \citep{Dhouib2022}. However, none of these theoretical predictions have been applied in practical applications to single stars, and their quality remains to be evaluated.

We have shown that the differences between the mean $\Delta P_{n}$ and/or p-mode frequencies $\nu_{\mathrm{p}}$ of a merger and genuine single star of similar mass are comfortably above the current best observational uncertainties. However, we have also shown in Fig.~\ref{fig:hrd_Pi1_shifts} that the merger product treated in this work will likely not manifest as an outlier based on its mean period spacing unless one has highly accurate $T_{\mathrm{eff}}$ and $L_{\star}$ measurements. Therefore, applications will benefit from additional constraints aside from asteroseismic ones. Surface diagnostics such as abundances and $L_{\star}/M_{\star}$ estimates, for example, could help to firmly identify MS merger products. Moreover, fitting observed properties with genuine single-star and merger models and their pulsation predictions may lead to systematic offsets in their derived masses, ages, pulsation frequencies, etc.\ between the two best solutions. For such forward modelling applications to become possible, extensive grids of merger product models are required in addition to single-star model grids.

It is imperative to keep in mind when interpreting the results presented in this work that we focus on one particular type of MS merger product, formed through the merger of two relatively young MS stars with a mass ratio close to one. The differences presented in this work are, therefore, lower limits on the differences that can be expected for merger products in general. From, for example, the set of merger products described in \citet{Glebbeek2013}, we know that depending on the age and binary configuration of a stellar merger's progenitor system, qualitatively different merger products can be achieved that might have different asteroseismic characteristics compared to equivalent genuine single stars. To answer the question of whether we can use asteroseismology to distinguish MS merger products from genuine single stars and to eventually create grids of MS merger models for forward modelling, the analysis presented in this work ought to be repeated on a range of different MS merger products. Doing this requires a cohesive set of 3D merger simulations, similar to those of \citet{Glebbeek2013}, but for mergers driven by binary evolution.  As demonstrated in this work, we should be wary of resorting to 1D merger prescriptions unless they are calibrated on the results of 3D simulations. Additionally, more realistic predictions for the asteroseismic fingerprints of stellar merger products also require us to take large-scale magnetic fields into account in our evolution models and asteroseismic predictions, given that they are predicted \citep{Schneider2019,Ryu2025} and inferred from observations \citep{Schneider2020,Frost2024} to be present in these stars.

\section*{Data availability}
The input and output files for this work are available at \href{https://doi.org/10.5281/zenodo.15194414}{https://doi.org/10.5281/zenodo.15194414}.

\begin{acknowledgements}
We wish to thank the referee, Nicholas Rui, for his valuable comments and suggestions, which have helped us further improve this work. We thank T.~Neumann for the valuable insights from her Bachelor thesis titled `Probing merger methods in a $9\,\msun$ and $8\,\msun$ binary star system' (University of Heidelberg, 2022). We thank L. Buchele, T. Van Reeth, Q. Copp\'ee, and D. Bowman (in no particular order) for the meaningful discussions and valuable comments and suggestions. Additional software used in this work includes \texttt{PyGYRE} \citep{Townsend2020pygyre}, \texttt{PyMesaReader} \citep{PyMesaReader2017}, \texttt{MPI for Python} \citep{Dalcin2005,Dalcin2021}, \texttt{Astropy} \citep{Astropy2013,Astropy2018,Astropy2022}, \texttt{NumPy} \citep{Harris2020}, and \texttt{SciPy} \citep{Scipy2020}. We used \texttt{Matplotlib} \citep{Hunter2007} for plotting. The authors acknowledge support from the Klaus Tschira Foundation. This work has received funding from the European Research Council (ERC) under the European Union’s Horizon 2020 research and innovation programme (Starting Grant agreement N$^\circ$ 945806: TEL-STARS, Consolidator Grant agreement N$^\circ$ 101000296: DipolarSound, and Synergy Grant agreement N$^\circ$101071505: 4D-STAR). While funded by the European Union, views and opinions expressed are however those of the author(s) only and do not necessarily reflect those of the European Union or the European Research Council. Neither the European Union nor the granting authority can be held responsible for them. This work is supported by the Deutsche Forschungsgemeinschaft (DFG, German Research Foundation) under Germany’s Excellence Strategy EXC 2181/1-390900948 (the Heidelberg STRUCTURES Excellence Cluster).
\end{acknowledgements}

\bibliographystyle{aa}
\bibliography{arXiv}

\clearpage

\onecolumn

\FloatBarrier

\begin{appendix}

\section{Additional comparisons between 3D MHD merger product and corresponding genuine single star structures}\label{app:envelope}
\begin{figure}[h!]
    \centering
    \begin{subfigure}{0.475\hsize}
        \centering
        \includegraphics[width=\linewidth]{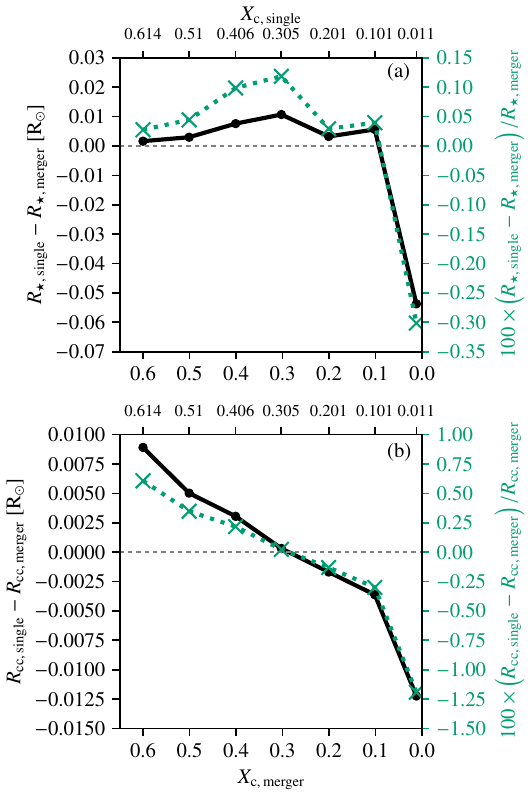}
        \caption{Absolute (left $y$-axis) and relative (right $y$-axis) differences in stellar radius $R_{\star}$ (panel a) and convective core radius $R_{\mathrm{cc}}$ (panel b) of the $16.9\,\msun$ 3D MHD merger product and the $17.4\,\msun$ genuine single star as a function of their respective central hydrogen mass fractions $X_{\mathrm{c}}$. The central hydrogen mass fraction for the merger product (genuine single star), $X_{\mathrm{c,\,merger}}$ ($X_{\mathrm{c,\,single}}$), is shown on the bottom (top) $x$-axis.}
        \label{fig:LRMcc_mhd}
    \end{subfigure}
    \hfill
    \begin{subfigure}{0.475\hsize}
        \centering
        \includegraphics[width=\linewidth]{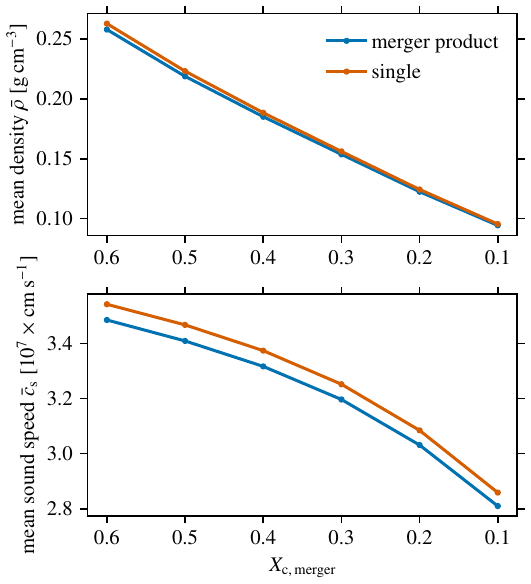}
        \caption{Mean envelope density (top panel) and sound speed (bottom panel) for the $16.9\,\msun$ 3D MHD merger product (blue line) and $17.4\,\msun$ genuine single star (red line) at different MS ages.}
        \label{fig:envelope_rho_cs}
    \end{subfigure}
    \caption{Comparison between the stellar and convective core radii (left) and mean envelope density and sound speed (right) of the $16.9\,\msun$ 3D MHD merger product and $17.4\,\msun$ genuine single star.}
    \label{fig:structure}
\end{figure}

In this appendix, we show additional comparisons between the structures of the $16.9\,\msun$ 3D MHD merger product and $17.4\,\msun$ genuine single star models. Figure~\ref{fig:LRMcc_mhd} shows the absolute and relative differences in the stellar and convective core radii of the two models throughout their MS evolution. In Fig.~\ref{fig:envelope_rho_cs}, we compare the mean values of the density $\rho$ and sound speed $c_{\mathrm{s}}$ in the respective envelopes of the merger product and genuine single star. The bottom of the envelope is taken as the location where the mass coordinate $m=M_{\mathrm{cc}}$. We interpolated the density and sound speed on a grid of equal-sized (in radius) cells to arrive at weighted means.

\clearpage

\section{Period spacing patterns for $\ell=2$ modes of 3D MHD merger product}\label{app:l2modes}
\begin{figure*}[h!]
\centering
  \includegraphics[width=16cm]{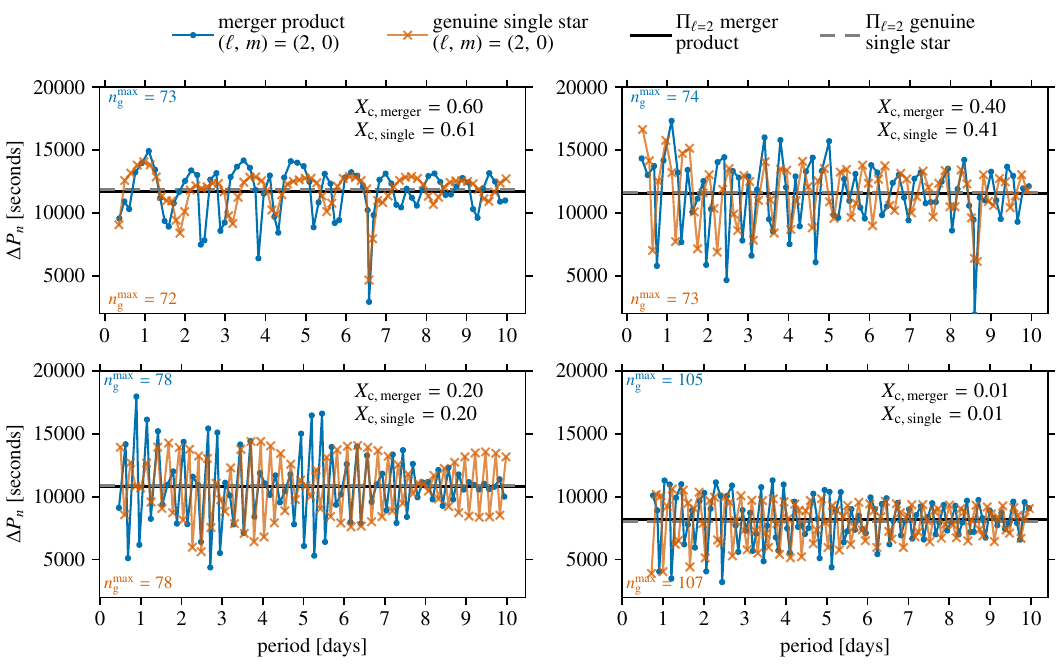}
     \caption{Same as Fig.~\ref{fig:psp_evolution}, now for $\ell = 2$ modes.}
     \label{fig:psp_evolution_l2}
\end{figure*}

Figure~\ref{fig:psp_evolution_l2} shows the PSPs for the $(\ell,\,m) = (2,\,0)$ modes of the $16.9\,\msun$ 3D MHD merger product and its corresponding $17.4\,\msun$ genuine single star.

\FloatBarrier

\section{Comparison of p modes with rotation}\label{app:rotating_pmodes}
\begin{figure}[h!]
    \centering
    \begin{subfigure}{0.475\hsize}
        \centering
        \includegraphics[width=\linewidth]{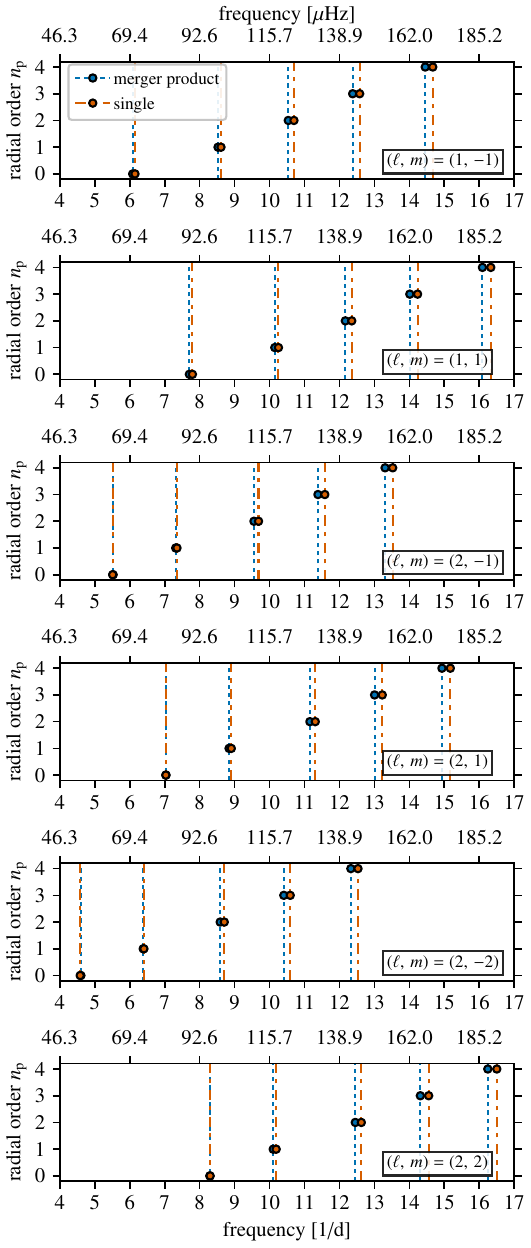}
        \caption{Frequencies and radial orders $n_{\mathrm{p}}$ of $(\ell,\,m) = (1,\,\pm 1)$, $(2,\,\pm 1)$, and $(2,\,\pm 2)$ p modes for the $16.9\,\msun$ 3D MHD merger product (solid lines) and the $17.4\,\msun$ genuine single star (dashed lines) at $X_{\mathrm{c}} = 0.50$ and with $\Omega/\Omega_{\mathrm{c}} = 0.30$. The dot markers indicate the radial order of each mode. The modes are shown in the inertial (observer's) frame.}
        \label{fig:pmodes_rot}
    \end{subfigure}
    \hfill
    \begin{subfigure}{0.475\hsize}
        \centering
        \includegraphics[width=\linewidth]{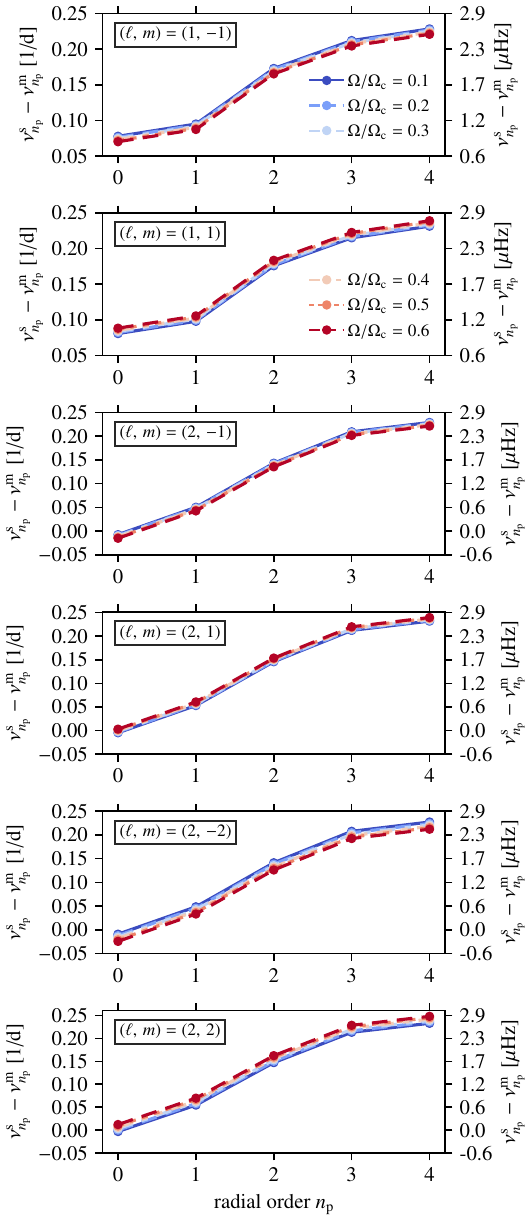}
        \caption{Absolute differences between the $17.4\,\msun$ genuine single star's p-mode frequencies $\nu_{n_\mathrm{p}}^{\mathrm{s}}$ and the $16.9\,\msun$ 3D MHD merger product's p-mode frequencies $\nu_{n_\mathrm{p}}^{\mathrm{m}}$ per radial order $n_\mathrm{p}$, with rotation rates of $\Omega/\Omega_{\mathrm{c}} = 0.10$, $0.20$, $0.30$, $0.40$, $0.50$, and $0.60$, at $X_{\mathrm{c}} = 0.50$. The lines are colour-coded according to the rotation rate. The modes are shown in the inertial (observer's) frame.}
        \label{fig:pmodes_rot_differences}
    \end{subfigure}
    \caption{Comparison between the predicted p-mode frequencies of the $16.9\,\msun$ 3D MHD merger product and $17.4\,\msun$ genuine single star with the inclusion of the effect of the Coriolis force.}
    \label{fig:pmodes_with_coriolis}
\end{figure}

In this appendix, we look at the effect of the Coriolis force on predicted p-mode frequencies. We treat the Coriolis force perturbatively and only up to first order in the rotation frequency, which is generally a good approximation for p modes in slow and modest rotators \citep{Aerts2019} and show the results of the inclusion of the Coriolis force for the models at $X_{\mathrm{c,\,merger}}=0.50$ in Figs.~\ref{fig:pmodes_rot} and \ref{fig:pmodes_rot_differences}. Figure \ref{fig:pmodes_rot} shows the mode frequencies and radial orders of $(\ell,\,m) = (1,\,\pm 1)$, $(2,\,\pm 1)$, and $(2,\,\pm 2)$ p modes predicted for the $16.9\,\msun$ 3D MHD merger product and $17.4\,\msun$ with $\Omega/\Omega_{\mathrm{c}} = 0.30$. As expected, prograde ($m>0$) p modes are shifted to higher frequencies, while retrograde ($m<0$) p modes are shifted to lower frequencies in the inertial frame. The zonal modes ($m=0$) are unaffected by rotation in the first-order Ledoux perturbative approach. Hence, we do not show them here again. Qualitatively, the frequency differences between the merger product and genuine single star look similar to those in the non-rotating case (Fig.~\ref{fig:pmodes_nonrot}). Quantitatively, we see from Fig.~\ref{fig:pmodes_rot_differences} that the differences between the p-mode frequencies of the same radial order are of the same order of magnitude as in the non-rotating case. Varying the rotation rate $\Omega/\Omega_{\mathrm{c}}$ does not affect the frequency differences between the merger product and genuine single star in a significant way. The effect of rotation is strongest for sectoral modes ($\ell=|m|$), where the differences increase with the rotation rate for prograde modes and decrease for retrograde modes.

\FloatBarrier
\section{Composition profiles and propagation diagrams for 1D merger methods}\label{app:profiles}
\begin{figure*}[h!]
\centering
  \includegraphics[width=16cm]{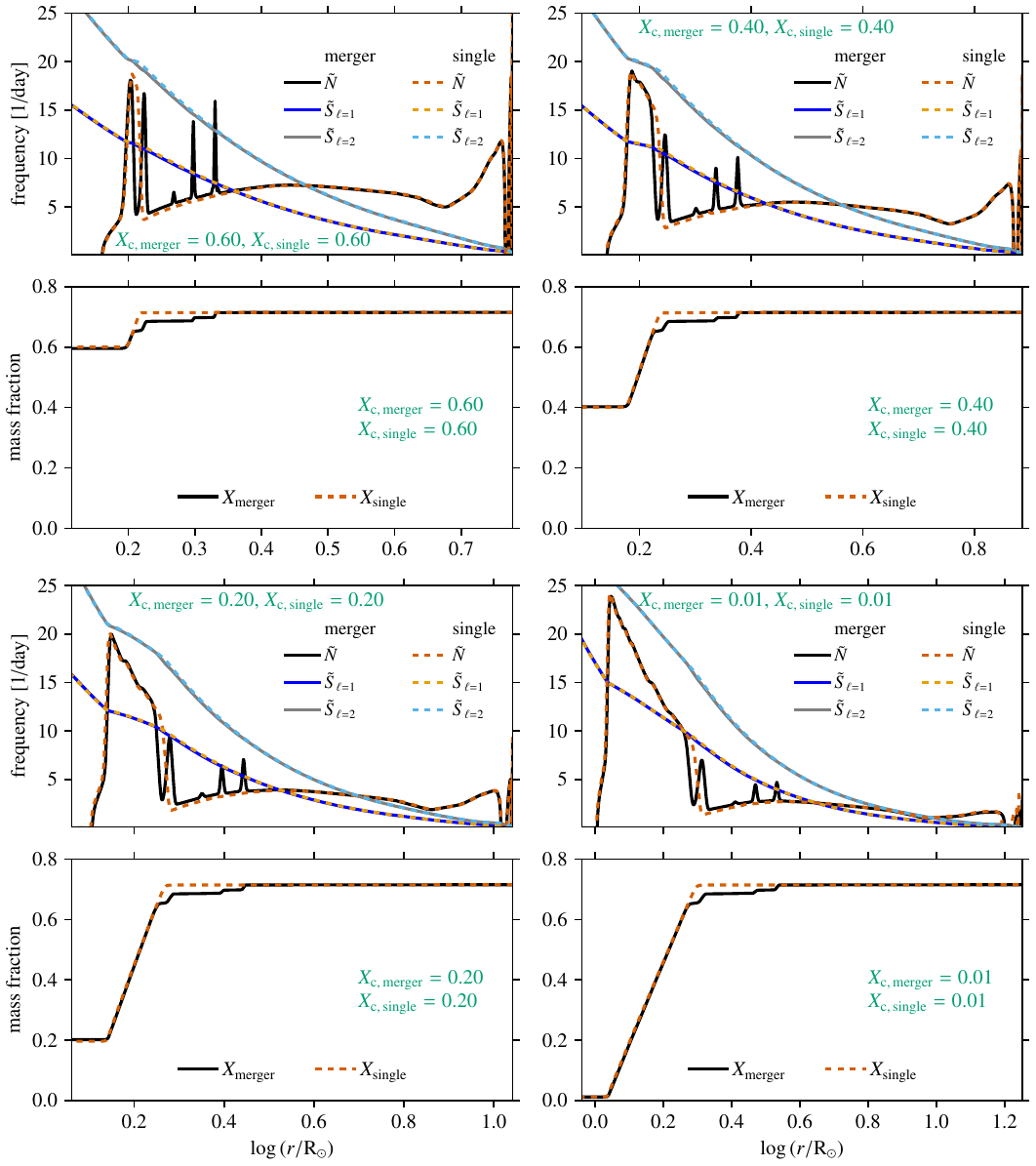}
     \caption{Same as Fig.~\ref{fig:mhd_props_and_comps_evolution}, now for the $16.9\,\msun$ entropy-sorted merger product (solid lines) and the $17.15\,\msun$ genuine single star (dashed lines).}
     \label{fig:es_props_and_comps_evolution}
\end{figure*}

\begin{figure*}[h!]
\centering
  \includegraphics[width=16cm]{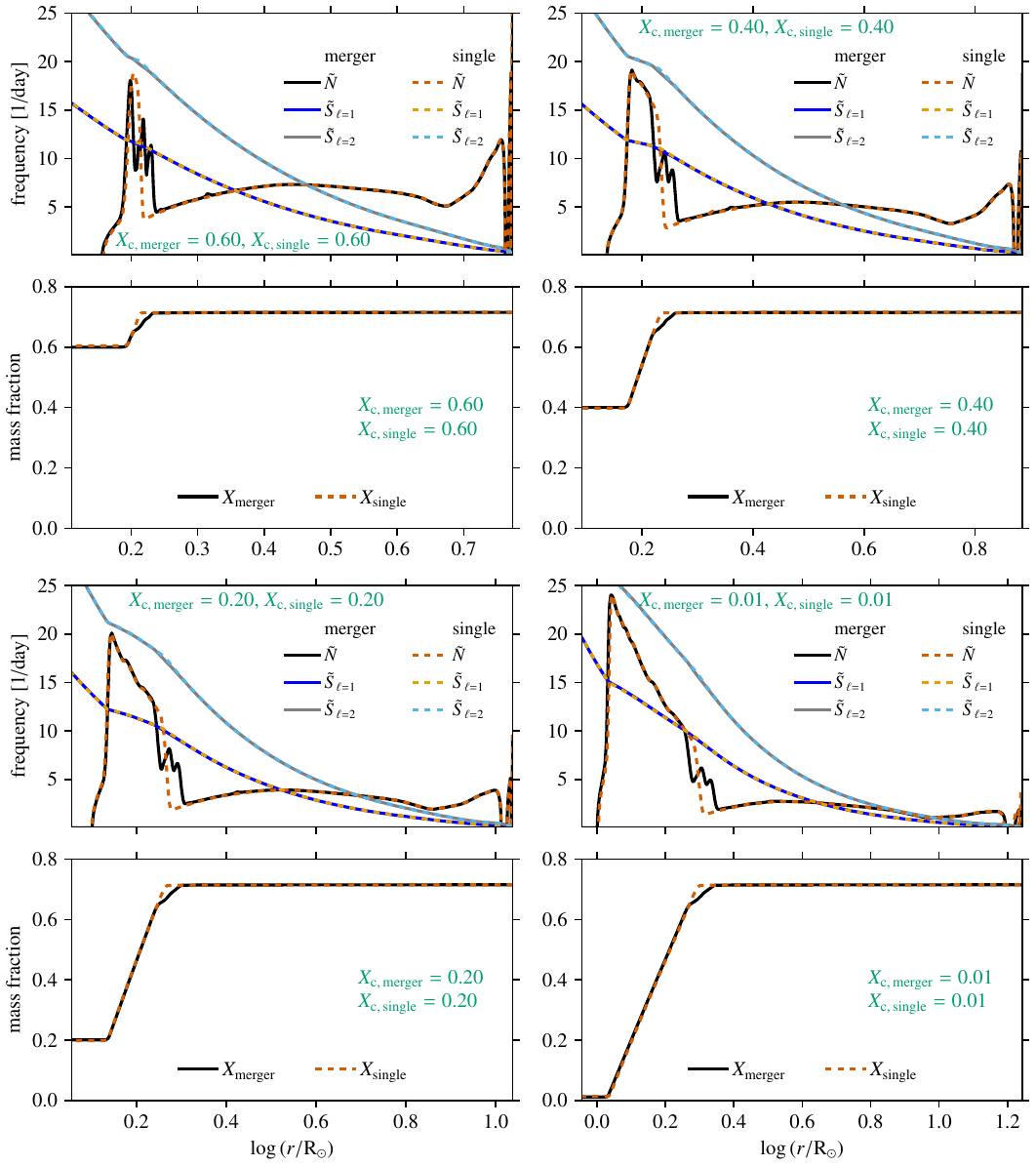}
     \caption{Same as Fig.~\ref{fig:mhd_props_and_comps_evolution}, now for the $16.9\,\msun$ \texttt{PyMMAMS} merger product (solid lines) and the $17.0\,\msun$ genuine single star (dashed lines).}
     \label{fig:pm_props_and_comps_evolution}
\end{figure*}

\begin{figure*}[h!]
\centering
  \includegraphics[width=16cm]{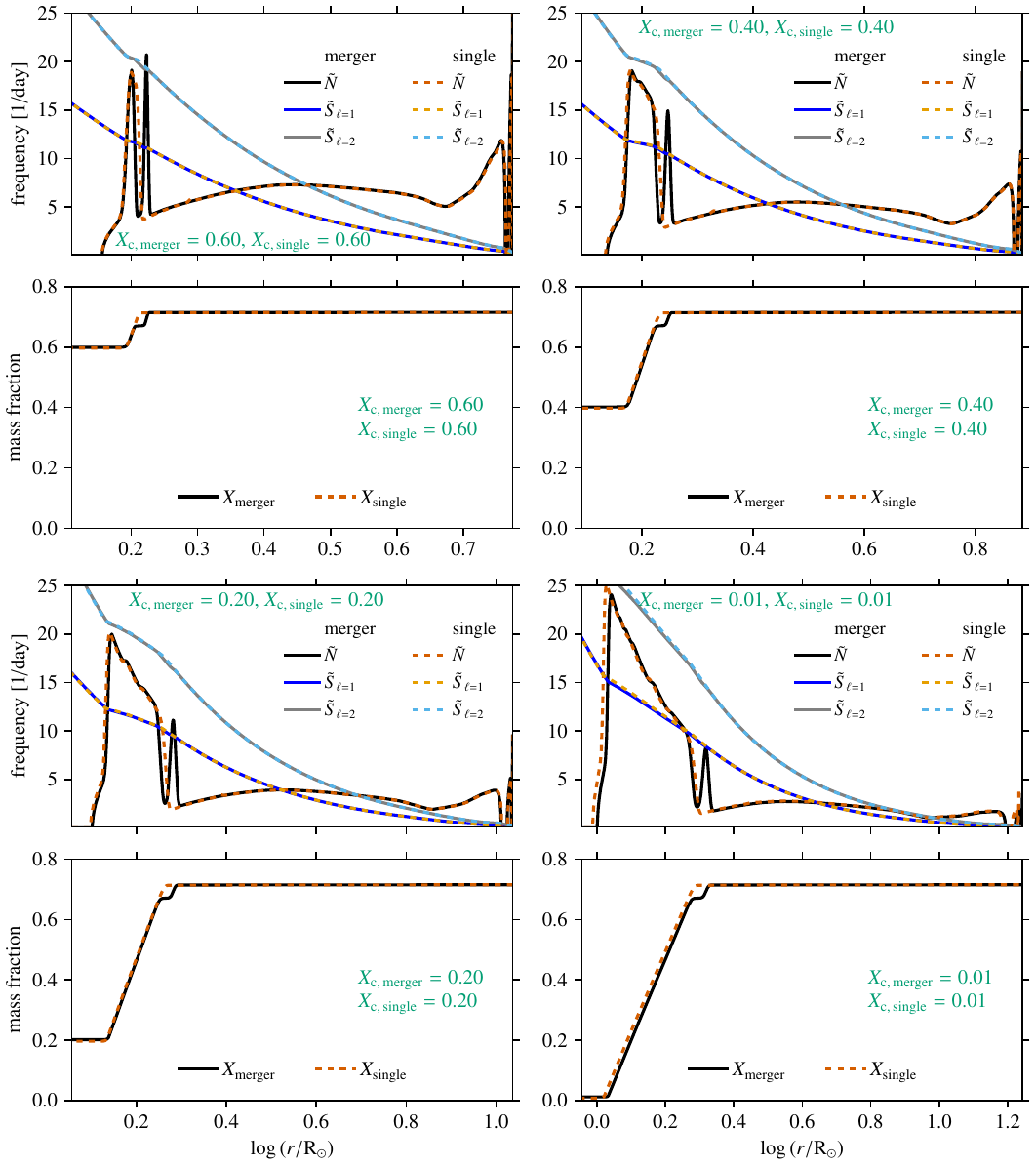}
     \caption{Same as Fig.~\ref{fig:mhd_props_and_comps_evolution}, now for the $16.9\,\msun$ fast accretion merger product (solid lines) and the $16.9\,\msun$ genuine single star (dashed lines).}
     \label{fig:fa_props_and_comps_evolution}
\end{figure*}

Figures \ref{fig:es_props_and_comps_evolution}--\ref{fig:fa_props_and_comps_evolution} show the evolution of the composition profiles and propagation diagrams with $X_{\mathrm{c}}$ for the $16.9\,\msun$ entropy-sorted, \texttt{PyMMAMS}, and fast accretion merger product models and their genuine single star counterparts.

\end{appendix}

\end{document}